\documentclass[final,3p,authoryear]{elsarticle}
\usepackage{amssymb}

\journal{Icarus}

\begin{document}
\begin{frontmatter}
\title{The Main-Belt Comets: The Pan-STARRS1 Perspective}
\author[asiaa,uh]{Henry H.\ Hsieh}\ead{hhsieh@asiaa.sinica.edu.tw}
\author[uh]{Larry Denneau}
\author[uh]{Richard J.\ Wainscoat}
\author[uh]{Norbert Sch\"orghofer}
\author[uh,upx]{Bryce Bolin}
\author[qub]{Alan Fitzsimmons}
\author[uh]{Robert Jedicke}
\author[uh]{Jan Kleyna}
\author[esrin]{Marco Micheli}
\author[uh]{Peter Vere{\v s}}
\author[uh]{Nicholas Kaiser}
\author[uh]{Kenneth C.\ Chambers}
\author[uh]{William S.\ Burgett}
\author[uh]{Heather Flewelling}
\author[uh]{Klaus W.\ Hodapp}
\author[uh]{Eugene A.\ Magnier}
\author[uh]{Jeffrey S.\ Morgan}
\author[princeton]{Paul A.\ Price}
\author[uh]{John L.\ Tonry}
\author[uh]{Christopher Waters}
\address[asiaa]{Institute of Astronomy and Astrophysics, Academia Sinica, P.O.\ Box 23-141, Taipei 10617, Taiwan}
\address[uh]{Institute for Astronomy, University of Hawaii, 2680 Woodlawn Drive, Honolulu, Hawaii 96822, USA}
\address[upx]{University of Phoenix, Hawaii Campus, 745 Fort Street, Honolulu, Hawaii 96813}
\address[qub]{Astrophysics Research Centre, Queens University Belfast, Belfast BT7 1NN, UK}
\address[esrin]{European Space Agency NEO Coordination Centre, Frascati, RM, Italy}
\address[princeton]{Department of Astrophysical Sciences, Princeton University, Princeton, NJ 08544, USA}

\begin{abstract}
We analyze a set of 760\,475 observations of 333\,026 unique main-belt objects obtained by the Pan-STARRS1 (PS1) survey telescope between 2012 May 20 and 2013 November 9, a period during which PS1 discovered two main-belt comets, P/2012 T1 (PANSTARRS) and P/2013 R3 (Catalina-PANSTARRS).
PS1 comet detection procedures currently consist of the comparison of the point spread functions (PSFs) of moving objects to those of reference stars, and the flagging of objects that show anomalously large radial PSF widths for human evaluation and possible observational follow-up.  Based on the number of missed discovery opportunities among comets discovered by other observers, we estimate an upper limit comet discovery efficiency rate of $\sim70$\% for PS1.  Additional analyses that could improve comet discovery yields in future surveys include linear PSF analysis, modeling of trailed stellar PSFs for comparison to trailed moving object PSFs, searches for azimuthally localized activity, comparison of point-source-optimized photometry to extended-source-optimized photometry, searches for photometric excesses in objects with known absolute magnitudes, and crowd-sourcing.
Analysis of the discovery statistics of the PS1 survey indicates an expected fraction of 59 MBCs per $10^6$ outer main-belt asteroids (corresponding to a total expected population of $\sim140$ MBCs among the outer main-belt asteroid population with absolute magnitudes of $12<H_V<19.5$), and a 95\% confidence upper limit of 96 MBCs per $10^6$ outer main-belt asteroids (corresponding to a total of $\sim$230 MBCs), assuming a detection efficiency of 50\%.  We note however that significantly more sensitive future surveys (particularly those utilizing larger aperture telescopes) could detect many more MBCs than estimated here.
Examination of the orbital element distribution of all known MBCs reveals an excess of high eccentricities ($0.1<e<0.3$) relative to the background asteroid population.  Theoretical calculations show that, given these eccentricities, the sublimation rate for a typical MBC is orders of magnitude larger at perihelion than at aphelion, providing a plausible physical explanation for the observed behavior of MBCs peaking in observed activity strength near perihelion.  These results indicate that the overall rate of mantle growth should be slow, consistent with observational evidence that MBC activity can be sustained over multiple orbit passages. [Accepted for publication in Icarus, 2014 Oct 19]
\end{abstract}


\end{frontmatter}

\section{INTRODUCTION}
\label{section:introduction}

\subsection{Active Asteroids: Main-Belt Comets and Disrupted Asteroids}
\label{section:mbcsdas}

In recent years, an increasing number of objects have been discovered that occupy asteroid-like orbits in the main asteroid belt but have shown evidence of comet-like activity, typically in the form of transient comet-like dust emission.  The suspected sources of this dust emission vary.  Some instances of activity are believed to result from comet-like sublimation of volatile sub-surface ice \citep[e.g.,][]{hsi04,hsi09a,hsi10b,hsi11b,hsi12b,hsi12c,hsi13a,mor11a,mor13,lic13a,jew14a,jew14b}, and the objects exhibiting this type of activity have come to be known as main-belt comets \citep[MBCs;][]{hsi06}.  For most MBCs, the presence of gas is only inferred by the presence and behavior of visible dust emission and is not directly detected.  However, a direct detection of water vapor outgassing has recently been made for main-belt object (1) Ceres by the $Herschel\ Space\ Observatory$ \citep{kup14}, marking the first time that sublimation on a main-belt object has been unambiguously detected.

In other instances, apparent comet-like dust emission is found to be the result of impacts, rotational destabilization, or a combination of several of these types of effects \citep[e.g.,][]{jew10,jew11b,jew13c,sno10,bod11,ish11a,ish11b,ste12b,mor11a,mor12,mor14}.  In these cases, the objects can be referred to as disrupted asteroids \citep[cf.][]{hsi12a}.  Instances where a combination of both sublimation and disruptive effects may be responsible for activity are also possible, such as the cases of 133P/Elst-Pizarro, for which rapid nucleus rotation may enhance the strength of its repeated dust emission events \citep{jew14b}, and P/2013 R3 (Catalina-PANSTARRS), for which rapid nucleus rotation may have induced its distintegration, but sublimation may have been responsible for ongoing post-disintegration activity \citep{jew14a}.  In such cases, the inferred presence of ice is the defining characteristic, and as such, we still consider these objects as MBCs.

MBCs have attracted interest in astrobiology for their potential to constrain theoretical studies indicating that material from the asteroid belt region could have been a significant primordial source of the water and other volatiles on Earth \citep[e.g.,][]{mor00,ray04,obr06,hsi14a}.  The existence of water in the asteroid belt in the past has long been inferred from the existence of hydrated minerals in CI and CM carbonaceous chondrite meteorites believed to originate from main-belt asteroids \citep[e.g.,][]{hir96,bur98,kei00}, as well as from spectroscopic observations of asteroids themselves \citep[cf.][]{riv02}. If ice is still present today in the asteroid belt, as the MBCs and other recent work \citep[e.g.,][]{riv10,cam10,tak12} suggests, it would represent a opportunity to probe a potential primordial water source through compositional and isotopic studies using either in situ measurements by a visiting spacecraft or some of the next-generation extremely large telescopes now in development.  Icy asteroids also contain some of the least altered material from the inner part of the protosolar disk still in existence today, and could give insights into the early stages of the formation of our solar system. The added bonus of their close proximity in the main asteroid belt means that in situ spacecraft studies are feasible given present-day technical capabilities.

\begin{figure}[h]
\centerline{\includegraphics[width=5.0in]{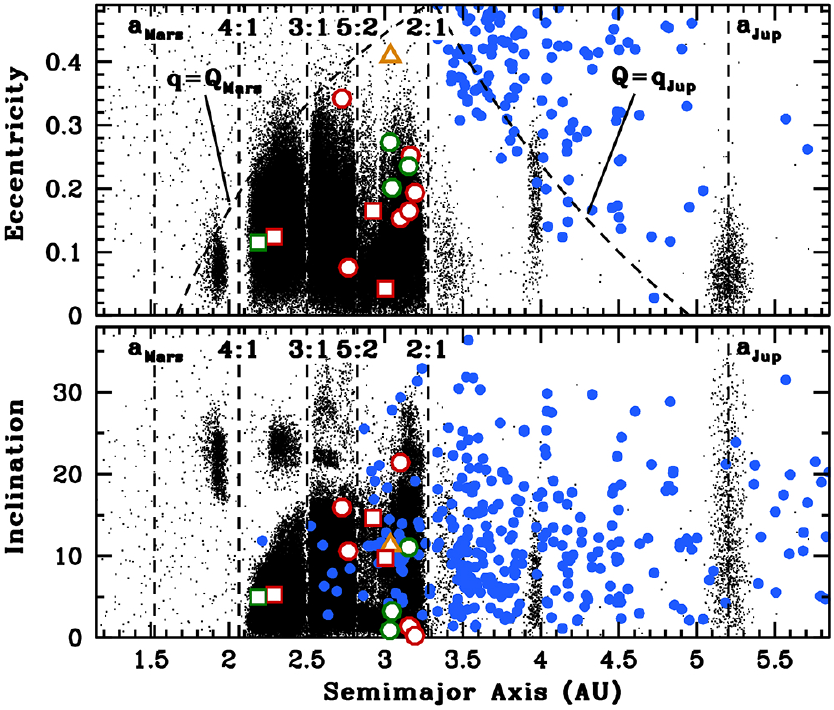}}
\caption{\small Plots of eccentricity (top) and inclination (bottom) vs. semimajor axis showing the distributions in orbital element space of main-belt asteroids (black dots), PS1-discovered MBCs (green circles), MBCs discovered by others (red circles), the one PS1-discovered disrupted asteroid (green square), disrupted asteroids discovered by others (red squares), and all other known comets (blue filled circles).  We also mark the orbital elements of 233P (orange triangle), which has not yet been classified as a MBC or disrupted asteroid.  Dotted lines mark the semimajor axes of Mars ($a_{\rm Mars}$) and Jupiter ($a_{\rm Jup}$), the semimajor axes of the 4:1, 3:1, 5:2, and 2:1 mean-motion resonances with Jupiter, and the loci of Mars-crossing orbits ($q=Q_{\rm Mars}$) and Jupiter-crossing orbits ($Q=q_{\rm Jup}$).
}
\label{figure:mbcs_aei}
\end{figure}

Collectively, MBCs and disrupted asteroids comprise the class of objects known as active asteroids \citep[e.g.,][]{jew12}.  Orbits of small solar system bodies are typically classified as asteroidal or cometary using the Tisserand parameter, or Tisserand invariant, with respect to Jupiter, $T_J$, as the primary dynamical discriminant, where asteroids have $T_J>3$ and comets have $T_J<3$ \citep{kre72}.  As such, a full accounting of active asteroids includes not only objects found in the main asteroid belt, but also other comet-like objects such as (2201) Oljato, (3200) Phaethon, and 107P/(4015) Wilson-Harrington \citep{rus84,jew13b,bow92}, which have $T_J>3$ but whose orbits carry them well outside the asteroid belt.  In this work here, however, we are primarily interested in objects in the main asteroid belt.  The currently known active asteroids found in the main asteroid belt, along with their orbital elements and absolute magnitudes, are listed in Table~\ref{table:knownaas}.  The most likely classification of each object as either a MBC or disrupted asteroid based on the available evidence \citep[e.g., numerical modeling indicating whether dust production is impulsive or ongoing, photometric measurements showing steady, increasing, or decreasing dust cross-sections, or observations of repeated activity; cf.][]{hsi12a} is also indicated, except for 233P/La Sagra, for which no physical analysis is available at this time.  We also plot the orbital elements of the known active asteroids in the main asteroid belt in Figure~\ref{figure:mbcs_aei}.

\setlength{\tabcolsep}{5pt}
\begin{table}[ht]
\caption{Known Active Main-Belt Asteroids$^{\dagger}$}
\smallskip
\scriptsize
\begin{tabular}{lcrrrrrrc}
\hline\hline
 \multicolumn{1}{c}{Name} &
 \multicolumn{1}{c}{Type$^a$} &
 \multicolumn{1}{c}{$a^b$} &
 \multicolumn{1}{c}{$e^c$} &
 \multicolumn{1}{c}{$i^d$} &
 \multicolumn{1}{c}{$T_J^e$} &
 \multicolumn{1}{c}{$P_{orb}^f$} &
 \multicolumn{1}{c}{$H_V^g$} &
 \multicolumn{1}{c}{Refs.$^h$} \\ 
\hline
(1) Ceres$^\ddagger$                       & MBC & 2.767 & 0.076 & 10.59 & 3.310 & 4.60 &     3.3 & [1]  \\
133P/Elst-Pizarro = (7968)$^\ddagger$      & MBC & 3.160 & 0.162 &  1.39 & 3.184 & 5.61 &    15.9 & [2]  \\
176P/LINEAR = (118401)$^\ddagger$          & MBC & 3.194 & 0.194 &  0.24 & 3.166 & 5.71 &    15.5 & [3]  \\
238P/Read                                 & MBC & 3.165 & 0.253 &  1.27 & 3.153 & 5.63 &    19.5 & [4]  \\
259P/Garradd                              & MBC & 2.726 & 0.342 & 15.90 & 3.217 & 4.50 &    20.1 & [5]  \\
288P/2006 VW$_{139}$ = (300163)$^\ddagger$ & MBC & 3.051 & 0.201 &  3.24 & 3.203 & 5.32 &    16.9 & [6]  \\
P/2010 R2 (La Sagra)                      & MBC & 3.099 & 0.154 & 21.40 & 3.099 & 5.46 &    18.8 & [7]  \\
P/2012 T1 (PANSTARRS)                     & MBC & 3.154 & 0.236 & 11.06 & 3.135 & 5.60 & $>$16.9 & [8]  \\
P/2013 R3 (Catalina-PANSTARRS)            & MBC & 3.033 & 0.273 &  0.90 & 3.184 & 5.28 & $>$15.4 & [9]  \\ 
(596) Scheila$^\ddagger$                   & DA  & 2.927 & 0.165 & 14.66 & 3.209 & 5.01 &     8.9 & [10] \\
P/2010 A2 (LINEAR)                        & DA  & 2.290 & 0.125 &  5.25 & 3.583 & 3.47 &    22.0 & [11] \\
P/2012 F5 (Gibbs)                         & DA  & 3.004 & 0.042 &  9.74 & 3.229 & 5.21 &    17.4 & [12] \\
P/2013 P5 (PANSTARRS)                     & DA  & 2.189 & 0.115 &  4.97 & 3.661 & 3.24 & $>$18.7 & [13] \\
233P/La Sagra                             & ?   & 3.037 & 0.409 & 11.28 & 3.081 & 5.29 & $>$18.6 & [14] \\ 
\hline
\hline
\end{tabular}
\newline {$^a$ Type of active asteroid (MBC: main-belt comet; DA: disrupted asteroid)}
\newline {$^b$ Osculating semimajor axis, in AU}
\newline {$^c$ Osculating eccentricity}
\newline {$^d$ Osculating inclination, in degrees}
\newline {$^e$ Tisserand parameter with respect to Jupiter}
\newline {$^f$ Orbital period, in years}
\newline {$^g$ Absolute $V$-band magnitude of nucleus}
\newline {$^h$ References:
[1] \citet{ted04,kup14}; 
[2] \citet{els96,hsi10b};
[3] \citet{hsi06a,hsi11a};
[4] \citet{rea05,hsi11b};
[5] \citet{gar08,mac12};
[6] \citet{hsi11c}; Hsieh et al., in prep;
[7] \citet{nom10,hsi14b};
[8] \citet{p2012t1,hsi13a};
[9] \citet{p2013r3,jew14a};
[10] \citet{ted04,lar10};
[11] \citet{bir10,jew10};
[12] \citet{gib12,nov14};
[13] \citet{p2013p5,jew13c};
[14] \citet{mai10}
}
\newline {$^\dagger$ All osculating orbital elements provided by JPL's online Small-Body Database Browser ({\tt http://ssd.jpl.nasa.gov/sbdb.cgi})}
\newline {$^\ddagger$ Previously known as an apparently inactive asteroid prior to discovery of comet-like activity}
\label{table:knownaas}
\end{table}

\subsection{Past Searches for MBCs}
\label{section:pastsearches}

Several past attempts have been made to find new active main-belt asteroids and constrain their abundance and spatial distribution in the asteroid belt.  First among these was the Hawaii Trails Survey \citep[HTP;][]{hsi09}, which led to the discovery of 176P/LINEAR \citep{hsi06a} and the subsequent recognition of MBCs as a new class of comets \citep{hsi06}.  As its objective was to find analogs of the first known MBC, comet 133P/Elst-Pizarro \citep{hsi04}, the HTP was designed as a targeted survey, focusing on members of asteroid families and small (km-scale) low-inclination main-belt asteroids.  It employed telescopes with a range of apertures (from 1~m up to 10~m), and covered a total of 599 asteroids.  Due to the relatively small number of objects and susceptibility of automated activity detection algorithms to false detections, screening for activity was largely done visually.  From this sample, a single MBC (176P) was discovered, establishing that 133P was not unique.  However, because the HTP was a targeted survey, the overall abundance of MBCs in the entire asteroid belt could not be explicitly calculated from its results.  It was nonetheless inferred that perhaps on the order of $\sim$100 MBCs could be present in the low-inclination, km-scale, outer-belt asteroid population (which consisted of $\sim10^4$ known objects at the time).

\citet{gil09} made the first systematic attempt to find MBCs using an untargeted survey, searching 12\,390 main-belt asteroids for activity using Canada-France-Hawaii Telescope (CFHT) Legacy Survey data, all obtained with the wide-field MegaCam imager on the 3.6~m CFHT.  Of that sample, 952 objects were subjected to detailed point spread function (PSF) analysis including comparisons of FWHM measurements of their PSFs to those of nearby field stars, visual comparison of their PSFs to stellar PSFs to look for excess flux in profile wings, and subtraction of stellar profiles from candidate object profiles to search for residual excess flux.  No objects were judged to exhibit real activity, with several false positive results attributed to nearby background objects, internal reflections, or other image artifacts.  Another 11\,615 objects in the survey were screened visually for activity, as in the HTP, and while the known active Centaur 166P/NEAT was detected in this second sample, no main-belt objects were found to exhibit cometary activity.  \citet{gil10} later published an update, reporting that another 13\,802 main-belt asteroids in CFHT Legacy Survey data had been visually examined, but none were found to exhibit cometary activity.  From these results, \citet{gil10} determined upper limits of currently active MBCs of $40\pm18$ in the entire asteroid belt, and $36\pm18$ in the outer main belt.

\citet{son11} placed upper limits on the number of MBCs using a sample of 924 main-belt asteroids observed by CFHT as part of the Thousand Asteroid Light Curve Survey \citep[TALCS;][]{mas09}.  Like the survey conducted by \citet{gil09}, TALCS was untargeted, and as such, was relatively unbiased in terms of the orbital element distribution of the objects observed by the survey, except in terms of inclination due to its focus on the ecliptic \citep{mas09}.  Two techniques were used to check for activity.  The first technique searched for directed tail-like emission by measuring relative fluxes in slices of an annulus around each object, while the second technique searched for deviations between object PSFs and modeled trailed stellar PSFs.  No MBCs were found using either technique.  Instead, the authors reported an upper limit ratio of MBCs to main-belt asteroids with absolute magnitudes of $H<21.0$ of $\sim$2500 MBCs for every $10^6$ main-belt asteroids, or $\sim$2000 MBCs for every $10^6$ main-belt asteroids in the outer belt alone.  The authors also found the intriguing result that about 5\% of the main-belt objects in the TALCS survey may exhibit low-level activity that cannot be identified for individual objects but is detectable as an aggregate statistical result.  They suggested that deeper observations of a large sample of main-belt objects may be able to confirm this result.

More recently, \citet{was13} reported on the search for comets in Palomar Transient Factory (PTF) data.  Approximately $2\times10^6$ Mould $R$-band or Gunn $g'$-band observations of $\sim$220\,000 main-belt objects obtained between 2009 March and 2012 July using Palomar Observatory's 1.2~m Oschin Schmidt Telescope were searched for cometary activity by calculating the ratio of each object's total flux within an elliptical aperture to its peak flux at its brightest pixel, and subtracting the median value of the same ratio computed for bright unsaturated stars in the same image.  The resulting difference indicated the degree of concentration of each object's flux relative to that of nearby field stars.  This parameter was used to identify 1577 comet candidates, which were then visually screened.  Most were found to be inactive objects that were contaminated by background sources or data artifacts, though genuine cometary activity was discovered in two non-main-belt objects that were previously labeled as asteroids: 2010 KG$_{43}$ and 2011 CR$_{42}$.  Two known MBCs, P/2010 R2 (La Sagra) and 288P/2006 VW$_{139}$, were also successfully identified as active.  From these results, \citet{was13} found a 95\% confidence upper limit of 33 active MBCs per $10^6$ main-belt asteroids assuming a detection efficiency of C=0.66 (their computed success rate of identifying known comets in their sample using the specified criteria), or 22 active MBCs per $10^6$ main-belt asteroids assuming a detection efficiency of C=1.00 (their success rate of identifying comets at least as extended as the known active MBCs in their sample).

Lastly, \citet{cik14} attempted to find active asteroids via photometry by searching the Minor Planet Center's (MPC) Observation Archive (``MPCAT-OBS'') for objects exhibiting significant deviations from their expected brightnesses.  Approximately $2.4\times10^6$ $V$-band photometric measurements of $\sim$300\,000 asteroids were analyzed and $\sim$1700 objects selected which had at least 5 measurements per object and showed $>3\sigma$ deviations between measured and predicted magnitudes (based on absolute magnitude and phase function parameters from the Minor Planet Center Orbit Database) for at least some of their observations.   From this sub-sample, six candidates (including 133P) were found to show such magnitude deviations over multiple nights.  Excluding 133P, the remaining five objects include three inner-main-belt objects (with semimajor axes of $2.064<a<2.501$~AU) and two middle-main-belt objects (with $2.501<a<2.824$~AU).  After investigating each of these candidates individually, the authors found that neither new observations that they obtained nor archival data for their candidate objects showed any evidence of visible resolved cometary activity.  In many cases, they found that they could not exclude factors such as contamination from nearby bright stars or changes in viewing geometry as alternative explanations for the observed brightness enhancements.  The authors noted that the low accuracy of the photometric data contained in the MPCAT-OBS archive was problematic for their study, but predicted that future surveys with more self-consistent photometry could use the technique to detect unresolved activity in known asteroids.

More generally, \citet{sol10} reported on a search for comets in Sloan Digital Sky Survey (SDSS) data using catalogued parameters measured by the SDSS photometric pipeline.  Using a training sample of serendipitously discovered comets found during visual inspection of SDSS images, they developed criteria for identifying comet candidates including a visual magnitude limit, a comparison between PSF and model magnitudes to identify resolved objects, and minimum sky-plane velocity cuts.  These criteria were used to identify 157\,996 candidates from the full SDSS DR5 catalog of 215 million objects.  Using even stricter PSF-model magnitude difference and sky-plane velocity cuts, the authors further reduced this sample to 43\,005 objects, which were then visually inspected.  The authors also tried applying color cuts (based on their initial training sample) to their 157\,996-object candidate sample, leaving them with 16\,254 color-selected candidates, which were then also visually inspected.  In the end, 19 comets were identified, or approximately one comet for every $10^7$ SDSS objects, though none were found to have main-belt orbits.

Of these past reported search attempts, only \citet{hsi09} discovered a previously unknown active asteroid, although \citet{was13} and \citet{cik14} were at least able to detect previously known active asteroids. Other active asteroids have generally been discovered serendipitously \citep[cf.][]{els96,rea05,gar08,bir10,nom10,lar10,mai10,gib12}.  In contrast, the active asteroid search efforts of the Pan-STARRS1 (PS1) survey have been substantially more effective, producing the discoveries of four new active asteroids to date \citep{hsi11c,p2012t1,p2013p5,p2013r3}, making PS1 currently the only observer or survey to have discovered multiple active asteroids.  In this paper, we discuss the details of the PS1 comet search effort, examine the statistical and physical implications of our results for studies of MBCs, and comment on possible approaches for future and ongoing active asteroid searches.

\section{OBSERVATIONS\label{observations}}

PS1 is a wide-field 1.8~m synoptic survey telescope located at the summit of Haleakala in Maui, Hawaii.  It employs a $3.2^{\circ}\times3.2^{\circ}$ 1.4 gigapixel camera consisting of 60 orthogonal transfer arrays, each comprising 64 590$\times$598 pixel CCDs.  PS1's filter system is modeled after that used by SDSS, and consists of filters designated $g_{\rm P1}$, $r_{\rm P1}$, $i_{\rm P1}$, $z_{\rm P1}$, $y_{\rm P1}$, and $w_{\rm P1}$ \citep{ton12}.  Of these, the $g_{\rm P1}$, $r_{\rm P1}$, $i_{\rm P1}$, and $z_{\rm P1}$ filters are similar to SDSS's $g'$, $r'$, $i'$, and $z'$ filters, while there is no corresponding SDSS filter for $y_{\rm P1}$, and the $w_{\rm P1}$ filter approximately spans the combined bandpasses of the $g_{\rm P1}$, $r_{\rm P1}$, and $i_{\rm P1}$ filters.  Of these, the $g_{P1}$, $r_{P1}$, $i_{P1}$, and $w_{P1}$ are the filters best-suited for observing solar system objects, and so for the analysis presented here, we will only consider observations made using these filters.

PS1 data are reduced using the system's image processing pipeline \citep[IPP;][]{mag06}.  In addition to standard data detrending, source extraction and transient identification (accomplished via pairwise-subtraction of successive images of individual fields), IPP also outputs various source characterization measurements including PSF moments, signal-to-noise, calibrated equivalent $V$-band magnitudes, and detection quality parameters.  The Moving Object Processing System \citep[MOPS;][]{den13} then searches these transient detections for known or suspected small solar system objects.  MOPS automatically identifies detections that are likely to be of the same object and links them together into a ``tracklet'' consisting of two or more detections made in a single night, and additionally identifies all tracklets that correspond to known objects.  Given the cadence and surveying strategy used by the PS1 telescope, typical asteroid or comet tracklets will consist of detections made all in the same filter ($g_{P1}$, $r_{P1}$, $i_{P1}$, or $w_{P1}$), or a combination of $g_{P1}$, $r_{P1}$, and $i_{P1}$ detections. Consecutive detections of an object belonging to a single individual tracklet are typically separated by approximately 10-15 minutes and have typical exposure times of 30-45 seconds.

PS1 has been operational since late 2008, and has been engaged in full survey operations since May 2010. As the survey has progressed though, data processing procedures and comet screening procedures have evolved, finally becoming mostly stable in May 2012.  For this analysis, we consider observations made between 2012 May 20 and 2013 Nov 09 in order to ensure approximately consistent screening rigor across our sample data set.
To ensure reliability of morphology parameter measurements across our data set, we also require median $S/N$ values of $5.0<S/N<125$.  These $S/N$ cuts are applied because IPP's morphology measurements are unreliable for detections that are either of poor quality or saturated.  IPP also provides various data quality metrics related to the amount of masking and number of bad or negative pixels in the vicinity of a detection.  Combining these metrics, we compute an internally-defined ``{\tt psfquality}'' parameter that ranges from 0 to 1, and require median {\tt psfquality} values of $>$0.4 for all tracklets (each of which we will hereafter consider to comprise a single ``observation'' of an object) considered in this analysis.

\begin{figure}
\centerline{\includegraphics[width=6.5in]{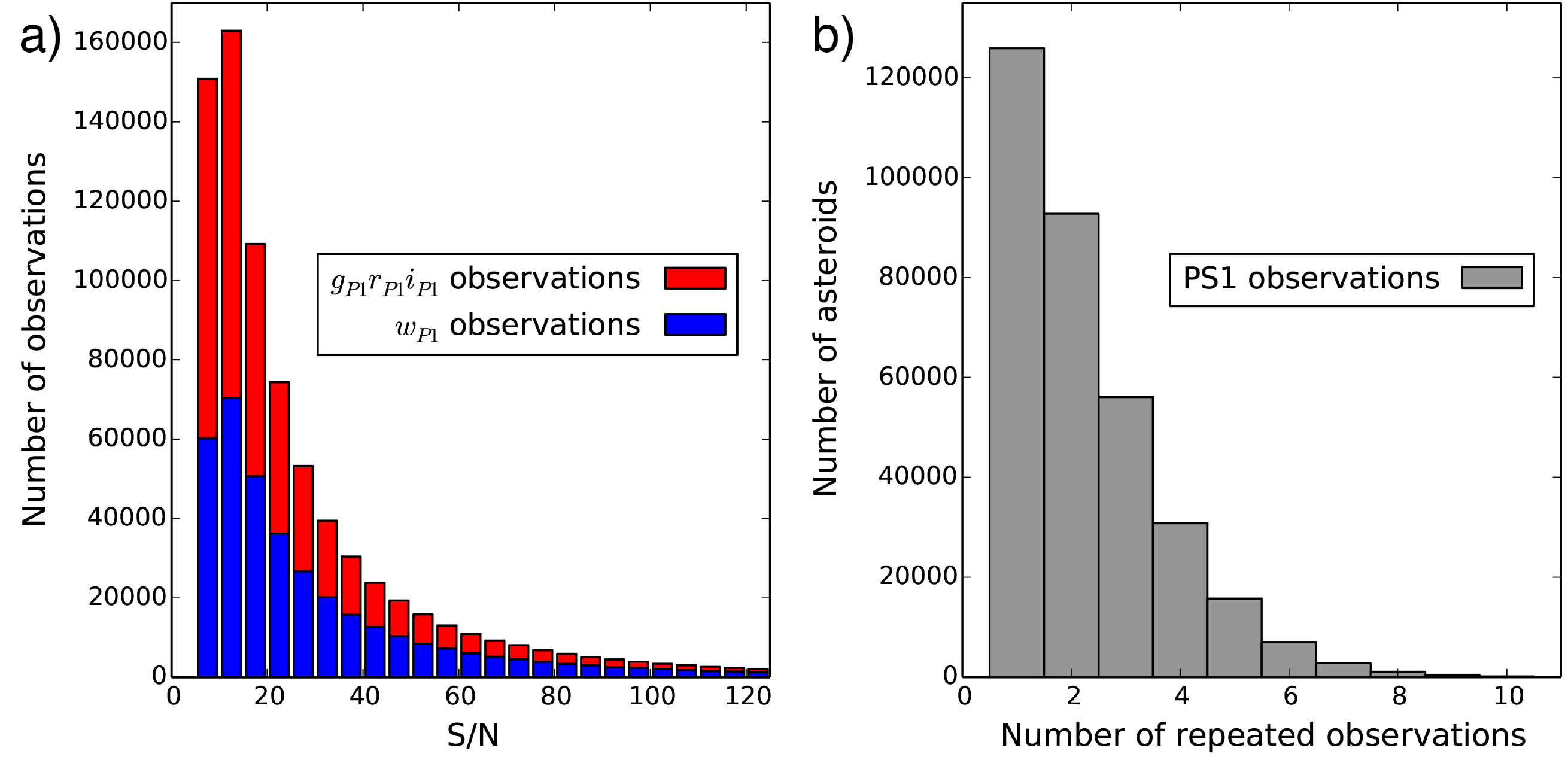}}
\caption{\small 
(a) Histogram of the measured median signal-to-noise (S/N) values of all observations made within the time period considered here and meeting other criteria described in the text.  Observations made using PS1's $g_{P1}$, $r_{P1}$, or $i_{P1}$ filters are indicated with red bars, and observations made using PS1's $w_{P1}$ filter are indicated by blue bars.
(b) Histogram of the number of repeated observations of the individual asteroids observed by PS1.
}
\label{figure:hist_s2n_filt_repeats}
\end{figure}

Finally, since we want to find active asteroids in the main asteroid belt, we only consider objects with main-belt orbits, i.e., with semimajor axes of $2.064~{\rm AU} < a < 3.277~{\rm AU}$, eccentricities of $e<0.45$, and inclinations of $i<40^{\circ}$.  Applying all of these criteria, and considering only $g_{\rm P1}$-, $r_{\rm P1}$-, $i_{\rm P1}$-, or $w_{\rm P1}$-band observations, we are left with a data set consisting of 760\,475 total observations of 333,026 unique asteroids (Table~\ref{table:sampledist}).  Of our individual observations, 56\% had $5<S/N<20$, 32\% had $20<S/N<50$, and 13\% had $50<S/N<125$ (Figure~\ref{figure:hist_s2n_filt_repeats}a).  Of the unique asteroids in our data sample, 38\% were observed once, 28\% were observed twice, 17\% were observed three times, and 17\% were observed four or more times (Figure~\ref{figure:hist_s2n_filt_repeats}b).


\begin{table}[ht]
\caption{Sample Distribution}
\footnotesize
\begin{tabular}{lcccccrr}
\hline\hline
 {} &
 {$a$ range$^a$} &
 {$e$ range$^b$} &
 {$i$ range$^c$} &
 {$H_V$ range$^d$} &
 {$m_V$ range$^e$} &
 {Unique Asts.$^f$} &
 {Total Obs.$^g$} \\
 \hline
Total            & $2.064-3.277$ & $e<0.45$ & $i<40^{\circ}$ & $8.3-23.6$ & $14.4-22.6$ & 333,026 & 760,475 \\
Inner main belt  & $2.064-2.501$ & $e<0.45$ & $i<40^{\circ}$ & $8.3-23.6$ & $14.4-22.5$ & 115,088 & 260,568 \\
Middle main belt & $2.501-2.824$ & $e<0.45$ & $i<40^{\circ}$ & $9.0-21.6$ & $14.5-22.5$ & 121,206 & 275,332 \\
Outer main belt  & $2.824-3.277$ & $e<0.45$ & $i<40^{\circ}$ & $9.6-20.1$ & $14.5-22.6$ &  96,732 & 224,575 \\
\hline
\hline
\end{tabular}
\newline {$^a$ Range of osculating semimajor axis distances, in AU}
\newline {$^b$ Range of osculating eccentricities}
\newline {$^c$ Range of osculating inclinations, in degrees}
\newline {$^d$ Range of absolute $V$-band magnitudes}
\newline {$^e$ Range of apparent $V$-band magnitudes at the times of observations}
\newline {$^f$ Number of unique asteroids observed}
\newline {$^g$ Number of total asteroid observations}
\label{table:sampledist}
\end{table}

\section{ANALYSIS AND RESULTS}
\label{section:results}

\subsection{Data Sample Characterization}
\label{section:samplecharacterization}

\begin{figure}
\centerline{\includegraphics[width=5.0in]{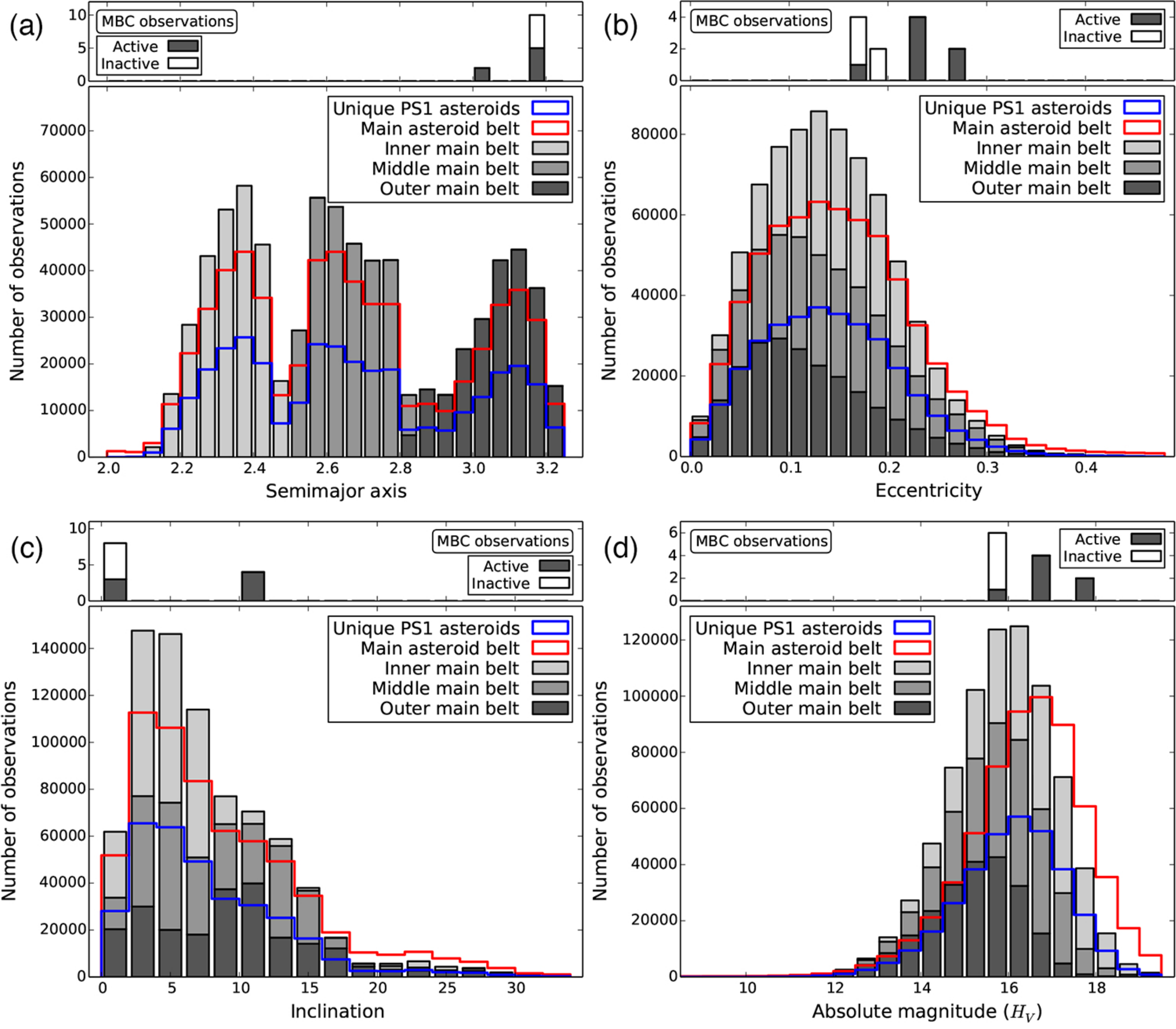}}
\caption{\small 
Histograms showing the (a) semimajor axis, (b) eccentricity, (c) inclination, and (d) absolute magnitude distributions of asteroids observed by PS1 during the time period considered here (grey bars), with histograms showing the distribution of all known asteroids overlaid as red lines.  Observations of asteroids from the inner, middle, and outer main belt (as defined in the text) are indicated with light, medium, and dark gray bars, respectively.
  }
\label{figure:hist_aeiH}
\end{figure}

While the PS1 survey is not a targeted survey like the HTP, it nonetheless has observational biases, many of which are shared with previous untargeted surveys (Section~\ref{section:introduction}).  In Figure~\ref{figure:hist_aeiH}, we show the distributions of the semimajor axes, eccentricities, inclinations, and absolute magnitudes of the asteroids observed in our data set.  In these plots, gray bars show the distributions of the orbital elements and absolute magnitudes associated with all of our asteroid observations (i.e., asteroids observed on multiple occasions will be represented multiple times in these distributions), while blue lines show the orbital element and absolute magnitude distributions of the observed population (i.e., only counting each asteroid once), and red lines show the orbital element and absolute magnitude distributions of the known asteroid population as determined from the asteroid orbital elements database (``{\tt astorb.dat}'') maintained at Lowell Observatory.

The shapes of the semimajor axis distributions of both our observed asteroid population and total observations are similar to that of the semimajor axis distribution of all known asteroids.  We note that the eccentricity distribution of the PS1-sampled asteroid population also follows the known population fairly well, though we see a drop-off in total observations of high-eccentricity objects.  We attribute this deficit to the fact that more eccentric asteroids spend less time near perihelion, where they are most easily observed, and more time along more distant portions of their orbits where smaller asteroids may be too faint to be detected by PS1.  We find a similar deficit of high-inclination asteroids in both our observed main-belt population as well as our total observations relative to the known main-belt population, likely due to the fact that the $w_{\rm P1}$ portion of our survey ($\sim$47\% of our total data set) mostly targeted areas of the sky near the ecliptic.  Finally, we note a deficit of objects with large absolute magnitudes (i.e., small sizes, low albedos, or both) for both our observed population and total observations, an indication of the bias of any untargeted observational survey against smaller, fainter objects that are more difficult to detect.

\begin{figure}
\centerline{\includegraphics[width=5.0in]{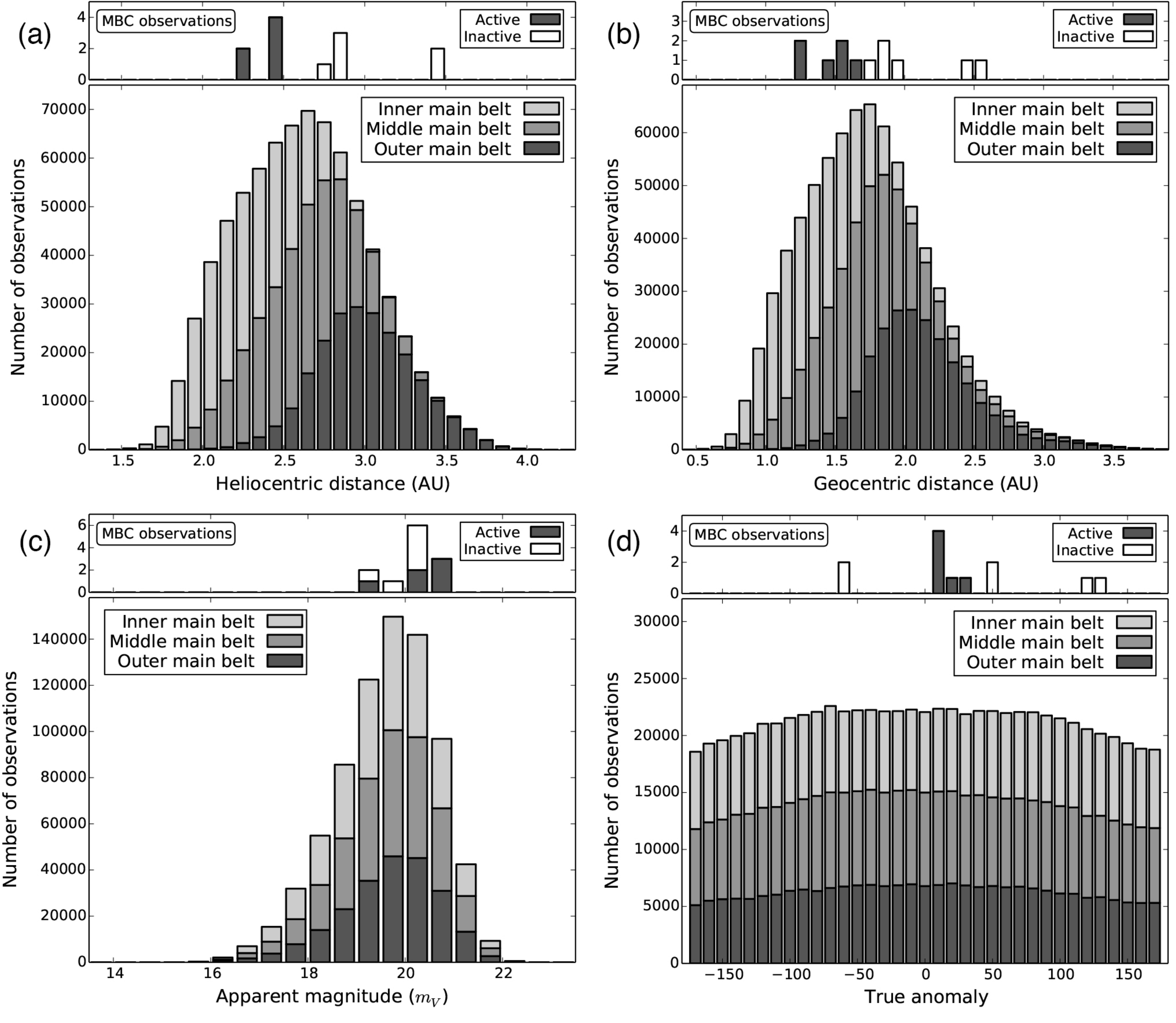}}
\caption{\small 
Histograms showing the (a) heliocentric distance, (b) geocentric distance, (c) apparent magnitude, and (d) true anomaly distributions of asteroids observed by PS1 during the time period considered here.  Observations of asteroids from the inner, middle, and outer main belt (as defined in the text) are indicated with light, medium, and dark gray bars, respectively.
  }
\label{figure:hist_rdmv}
\end{figure}

In Figure~\ref{figure:hist_rdmv}, we show the distributions of the heliocentric and geocentric distances, apparent magnitudes, and true anomalies of all asteroids in our data set at the times of their observations.  As might be expected, we find slight skewings of the distributions towards small heliocentric distances, small geocentric distances, and true anomalies closer to perihelion, since these conditions contribute to asteroids having larger apparent magnitudes, and therefore being more easily detected.  We furthermore find a peak in our observed apparent magnitude distribution at $m_V\sim20$~mag and a magnitude cut-off at the faint end of $m_V\sim22$~mag, beyond which nearly all of observations we considered failed to meet our $S/N>5.0$ data quality requirement.

\subsection{Comet Screening Procedures}
\label{section:ps1comets}

In practice, the procedure for searching for active asteroids is identical to that for searching for comets in general.  Therefore, if we simply strive to detect as many comet-like objects in our data as possible, some fraction of our discoveries will be active asteroids.
As described in \citet{hsi12b}, potential comets in PS1 data are identified using a ``{\tt psfextent}'' parameter, given by
\begin{equation}
{\tt psfextent} = \left({{\tt moments\_xx}\times{\tt moments\_yy}\over{\tt psf\_major}\times{\tt psf\_minor}}\right)^{1/2}
\end{equation}
where {\tt moments\_xx} and {\tt moments\_yy} are the measured second linear PSF moments of each transient source in the $x$ and $y$ directions, respectively.  The parameters {\tt psf\_major} and {\tt psf\_minor} are the expected half-widths at half-maximum of each source's PSF along the directions of its major and minor axes, respectively, where the expected PSF is assumed to be elliptical in shape.  This expected PSF is modeled using high-significance detections identified as point sources from across the PS1 camera's field-of-view, and varies as a function of spatial position in the field.  The {\tt psfextent} parameter essentially characterizes how extended a source's PSF is compared to nearby stars and can indicate how likely an object is to be cometary.

\begin{figure}
\centerline{\includegraphics[width=6in]{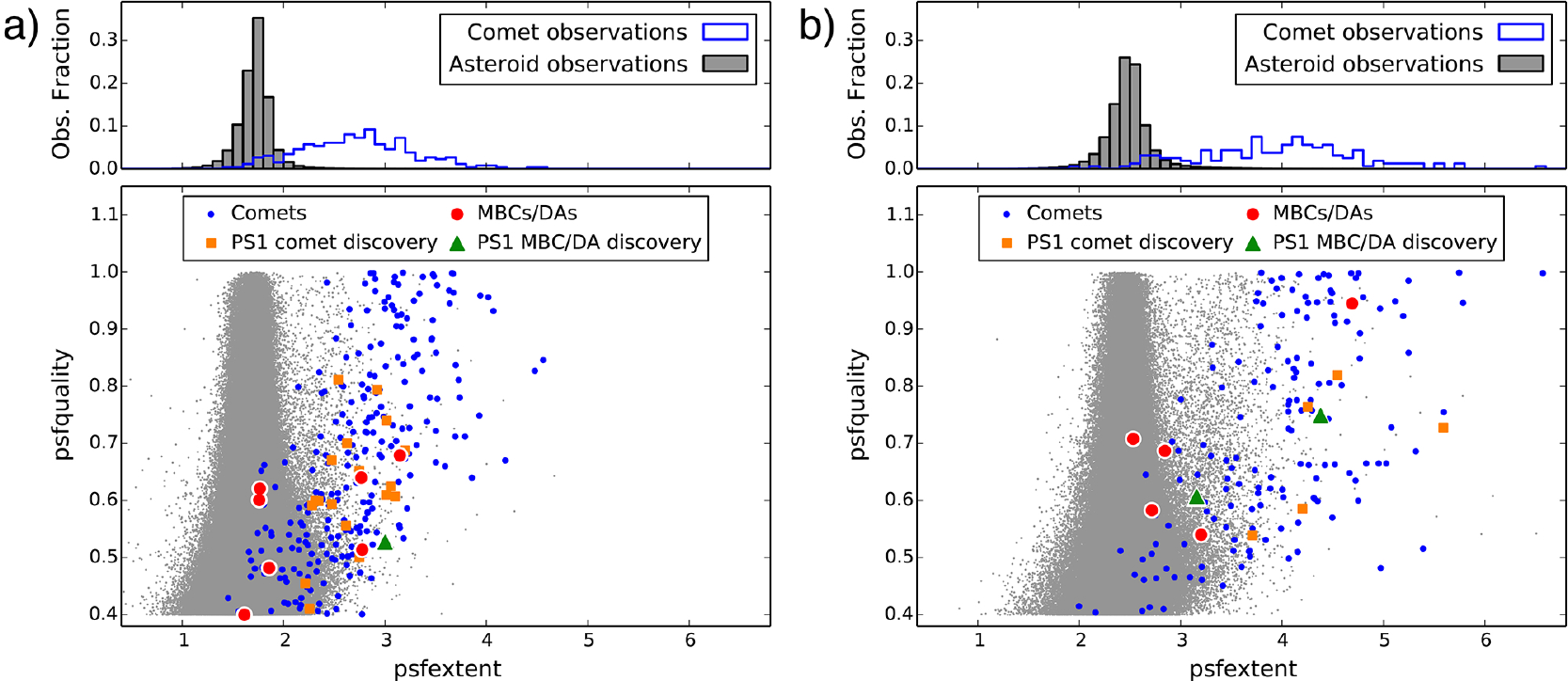}}
\caption{\small (a) Plot (lower panel) showing the distribution of {\tt psfextent} and {\tt psfquality} values for all asteroid (small gray dots) and comet (blue circles) observations in our data sample that were obtained prior to 2013 July 13, where discovery observations of PS1-discovered comets are marked with orange squares.  PS1 observations of active asteroids are also marked, with green triangles representing discovery observations of PS1-discovered active asteroids, and red circles representing non-discovery observations.  An accompanying histogram (top panel) shows the normalized {\tt psfextent} distributions of the asteroid (gray bars) and comet observations (blue line) in our data. (b) Same as (a) but for observations obtained after 2013 July 13.
}
\label{figure:hist_fuzz_psfq}
\end{figure}

To illustrate the effectiveness of {\tt psfextent} as a cometary discriminant, we plot median {\tt psfextent} values against median {\tt psfquality} values for all asteroid observations in our data set in Figure~\ref{figure:hist_fuzz_psfq}\footnote{Due to a change in the way IPP measures PSF moments on 2013 July 13, the {\tt psfextent} value measured for a particular detection changed by a factor of $\sqrt{2}$ on that date, and so for clarity, we construct separate plots for data taken before and after that date.}.  We then over-plot the same parameters for comet observations (for comets known as of 2014 February 8) obtained during the same time period.  This data set consists of 412 observations of 176 unique comets (though not all of the comets were active when observed by PS1), and 28 were new discoveries (Table~\ref{table:ps1discoveries}).  

\setlength{\tabcolsep}{2.5pt}
\begin{table}[ht]
\caption{PS1 Comet Discoveries (2012 May 20 - 2013 Nov 9)}
\smallskip
\scriptsize
\begin{tabular}{llrrrrcrrcc}
\hline\hline
 {} &
 {Disc.\ Date$^a$} &
 \multicolumn{1}{c}{$q^b$} &
 \multicolumn{1}{c}{$e^c$} &
 \multicolumn{1}{c}{$i^d$} &
 \multicolumn{1}{c}{$T_J^e$} &
 \multicolumn{1}{c}{$R^f$} &
 \multicolumn{1}{c}{$\nu^g$} &
 \multicolumn{1}{c}{$m_V^h$} &
 \multicolumn{1}{c}{Type$^i$} &
 \multicolumn{1}{c}{Ref.$^j$} \\
 \hline
C/2012 S3 (PANSTARRS)                   & 2012 Sep 27 & 2.308 & 1.001 & 112.9 & $-$0.729 & 4.245 & 275.0 & 20.2 & LPC     & [1] \\ 
C/2012 S4 (PANSTARRS)                   & 2012 Sep 28 & 4.349 & 1.000 & 126.5 & $-$1.540 & 4.884 & 321.3 & 19.3 & LPC     & [2] \\ 
P/2012 T1 (PANSTARRS)                   & 2012 Oct  6 & 2.411 & 0.236 &  11.1 &    3.135 & 2.415 &   7.5 & 20.6 & MBC     & [3] \\ 
P/2012 SB$_{6}$ (Lemmon)                & 2012 Oct  8 & 2.406 & 0.385 &  11.0 &    2.902 & 2.412 & 352.7 & 19.3 & JFC     & [4] \\ 
P/2012 TK$_{8}$ (Tenagra)               & 2012 Oct  9 & 3.091 & 0.261 &   6.3 &    2.964 & 3.264 & 318.1 & 20.7 & JFC     & [5] \\ 
271P/van Houten-Lemmon (2012 TB$_{36}$) & 2012 Oct  9 & 4.256 & 0.388 &   6.9 &    2.864 & 4.471 & 325.5 & 21.0 & JFC     & [6] \\ 
P/2012 T2 (PANSTARRS)                   & 2012 Oct 10 & 4.821 & 0.160 &  12.6 &    2.930 & 4.855 & 341.6 & 21.1 & JFC     & [7] \\ 
P/2012 T3 (PANSTARRS)                   & 2012 Oct 10 & 2.287 & 0.657 &   9.6 &    2.464 & 2.783 &  56.6 & 21.4 & JFC     & [8] \\ 
C/2012 U1 (PANSTARRS)                   & 2012 Oct 18 & 5.265 & 0.999 &  56.3 &    1.577 & 6.959 & 300.9 & 21.0 & LPC     & [9] \\ 
P/2012 U2 (PANSTARRS)                   & 2012 Oct 21 & 3.628 & 0.507 &  10.5 &    2.723 & 3.639 & 352.3 & 20.5 & JFC     & [10] \\ 
C/2012 V1 (PANSTARRS)                   & 2012 Nov  3 & 2.089 & 0.999 & 157.8 & $-$1.658 & 3.584 & 279.5 & 20.4 & LPC     & [11] \\ 
C/2012 X2 (PANSTARRS)                   & 2012 Dec 12 & 4.748 & 0.771 &  34.1 &    2.356 & 4.807 & 346.4 & 19.9 & JFC     & [12] \\ 
P/2012 WA$_{34}$ (Lemmon-PANSTARRS)     & 2013 Jan  7 & 3.173 & 0.340 &   6.12 &   2.880 & 3.175 & 356.4 & 21.2 & JFC     & [13] \\ 
281P/MOSS (2013 CE$_{31}$)              & 2013 Feb  9 & 4.018 & 0.174 &   4.72 &   2.968 & 4.126 &  35.3 & 21.0 & JFC     & [14] \\ 
C/2013 G3 (PANSTARRS)                   & 2013 Apr 10 & 3.852 & 1.000 &  64.68 &   1.977 & 6.200 & 284.0 & 20.8 & LPC     & [15] \\ 
P/2013 G4 (PANSTARRS)                   & 2013 Apr 12 & 2.703 & 0.380 &   6.04 &   2.848 & 2.647 &  17.0 & 20.9 & JFC     & [16] \\ 
C/2013 G8 (PANSTARRS)                   & 2013 Apr 14 & 5.141 & 0.999 &  27.62 &   2.492 & 5.390 & 335.2 & 20.0 & LPC     & [17] \\ 
P/2013 J4 (PANSTARRS)                   & 2013 May  5 & 2.287 & 0.646 &   4.76 &   2.500 & 2.402 & 331.5 & 21.0 & JFC     & [18] \\ 
P/2013 CU$_{129}$ (PANSTARRS)           & 2013 Jun  2 & 0.800 & 0.722 &  12.15 &   2.813 & 1.278 & 276.2 & 20.9 & JFC     & [19] \\ 
282P/(323137) 2003 BM$_{80}$            & 2013 Jun 12 & 3.451 & 0.188 &   5.81 &   2.990 & 3.501 &  24.7 & 20.1 & JFC     & [20] \\ 
P/2013 N3 (PANSTARRS)                   & 2013 Jul  4 & 3.029 & 0.592 &   2.17 &   2.624 & 3.453 & 312.1 & 20.7 & JFC     & [21] \\ 
P/2013 N5 (PANSTARRS)                   & 2013 Jul 14 & 1.823 & 0.732 &  23.24 &   2.197 & 1.828 &   6.4 & 21.4 & JFC     & [22] \\ 
P/2013 O2 (PANSTARRS)                   & 2013 Jul 16 & 2.146 & 0.439 &  13.31 &   2.860 & 2.438 & 307.4 & 20.6 & JFC     & [23] \\ 
P/2013 P1 (PANSTARRS)                   & 2013 Aug  1 & 3.390 & 0.605 &  18.70 &   2.543 & 3.573 &  30.2 & 20.1 & JFC     & [24] \\ 
C/2013 P4 (PANSTARRS)                   & 2013 Aug 15 & 5.967 & 0.597 &   4.26 &   3.051 & 6.278 & 330.1 & 20.8 & Centaur & [25] \\ 
P/2013 P5 (PANSTARRS)                   & 2013 Aug 15 & 1.936 & 0.115 &   4.97 &   3.661 & 2.147 & 272.8 & 21.0 & DA      & [26] \\ 
P/2013 R3 (Catalina-PANSTARRS)          & 2013 Sep 15 & 2.204 & 0.273 &   0.90 &   3.184 & 2.218 &  14.0 & 20.5 & MBC     & [27] \\ 
P/2013 T1 (PANSTARRS)                   & 2013 Oct  5 & 2.210 & 0.623 &  24.21 &   2.402 & 2.291 &  24.7 & 21.6 & JFC     & [28] \\ 
\hline
\hline
\end{tabular}
\newline {$^a$ UT date of discovery}
\newline {$^b$ Osculating perihelion distance, in AU}
\newline {$^c$ Osculating eccentricity}
\newline {$^d$ Osculating inclination, in degrees}
\newline {$^e$ Tisserand parameter with respect to Jupiter}
\newline {$^f$ Heliocentric distance at the time of observations, in AU}
\newline {$^g$ True anomaly, in degrees}
\newline {$^h$ Equivalent apparent $V$-band magnitude of nucleus}
\newline {$^i$ Type of cometary object (JFC: Jupiter-family comet; LPC: long-period comet; MBC: main-belt comet; DA: disrupted asteroid)}
\newline {$^j$ References:
[1] \citet{c2012s3},
[2] \citet{c2012s4},
[3] \citet{p2012t1},
[4] \citet{p2012sb6},
[5] \citet{p2012tk8},
[6] \citet{p2012tb36},
[7] \citet{p2012t2},
[8] \citet{p2012t3},
[9] \citet{c2012u1},
[10] \citet{p2012u2},
[11] \citet{c2012v1},
[12] \citet{c2012x2}
[13] \citet{p2012wa34},
[14] \citet{p2013ce31},
[15] \citet{c2013g3},
[16] \citet{p2013g4},
[17] \citet{c2013g8},
[18] \citet{p2013j4},
[19] \citet{p2013cu129},
[20] \citet{p2013bm80},
[21] \citet{p2013n3},
[22] \citet{p2013n5},
[23] \citet{p2013o2},
[24] \citet{p2013p1},
[25] \citet{c2013p4},
[26] \citet{p2013p5},
[27] \citet{p2013r3},
[28] \citet{p2013t1}
}
\label{table:ps1discoveries}
\end{table}
\setlength{\tabcolsep}{5pt}

Like in the comet screening plots shown in \citet{hsi12b}, point sources (i.e., inactive asteroids) in Figure~\ref{figure:hist_fuzz_psfq} cluster near a particular range of {\tt psfextent} values ($\sim1.5-1.9$ for data obtained prior to 2013 July 13, and $\sim2.3-2.7$ for data obtained after 2013 July 13), while active comets have larger {\tt psfextent} values.  The observed clustering of {\tt psfextent} values over a range of values rather than a single value is due to the non-sidereal motion of solar system objects, which causes slight amounts of trailing over the course of typical PS1 exposure, resulting in slightly elongated PSFs and larger PSF moment measurements than would be obtained for stationary, stellar sources.  Of course, this means that comet-like {\tt psfextent} values could be measured for very fast-moving objects, even in the absence of activity.  Fortunately, in practice, very few extremely fast-moving objects (usually near-Earth objects, or NEOs) are observed each night (relative to the more abundant observations of slow-moving main-belt objects, or MBOs), and can therefore be easily visually identified and disregarded, or can also be screened out by a maximum sky-plane velocity cut.

A broadening of the distribution of {\tt psfextent} values also occurs with decreasing {\tt psfquality} values, demonstrating the deterioration in the reliability of PSF moment measurements by IPP for poor-quality detections.  In practice, this means that larger {\tt psfextent} values are required for low-quality detections to be considered reliable cometary candidates: {\tt psfextent} values cannot be considered in isolation. In the PS1 comet screening process currently in place, we required that sources have {\tt psfquality}$>$0.4 in order to be flagged for visual inspection, more detailed PSF analysis and stacking, and possible observational follow-up.

\subsection{PS1 Detection Efficiency}
\label{section:efficiency}

For an active comet to be discovered in PS1 data, many stages of processing and screening must be successfully completed, all of which affect our overall comet detection rate.  We discuss these stages (including tracklet linking, flagging of possible cometary activity, visual screening, and follow-up observations) in detail in \ref{appendix:comet_screening}.  In summary, though, the PS1 comet discovery process is complex and includes several opportunities for comets to be missed or lost.  Many of the complexities and inefficiencies are difficult to quantify or simulate, particularly given the limited human power on the MOPS team for whom keeping up with the nightly flow of real data is already challenging enough without having to also process simulated data as is commonly done to measure efficiency in other surveys.

To get a rough estimate of our discovery efficiency, however, we note that PS observed 12 of the 65 comets discovered by others during the time period considered here prior to their reported discoveries, but did not note them as comets due to one or more of the reasons discussed in \ref{appendix:comet_screening}.  Accounting for these 12 missed discovery opportunities, PS1's 28 successful discovered comets during this time period then corresponds to a 0.70 efficiency rate (i.e., 28 successful discoveries out of 40 total opportunities).  This rate assumes however that other observers provide a complete accounting of all of the available new comets that PS1 misses, which is almost certainly not the case.  Given the multiple examples of PS1 identifying activity in objects shortly (e.g., a few days) after they were reported as asteroidal by their discoverers (e.g., P/2012 TK$_{8}$, 271P, 281P; Table~\ref{table:ps1discoveries}), it is apparent that PS1 is capable of discovering weakly active comets that others cannot.  As such, there was very likely a number of observable weakly active comets that were missed by PS1 due to human or instrumental factors, but were also unaccounted for by other observers who did not detect them due to insufficient sensitivity.  As such, the 70\% efficiency rate estimate computed here should be considered an upper limit, and we suggest that future data mining of the PS1 archive could uncover additional previously unknown comets, and perhaps even some new MBCs or other active asteroids.

\subsection{Main-Belt Comets \& Disrupted Asteroids in PS1 Data}
\label{section:ps1mbcs}

\begin{figure}
\centerline{\includegraphics[width=5in]{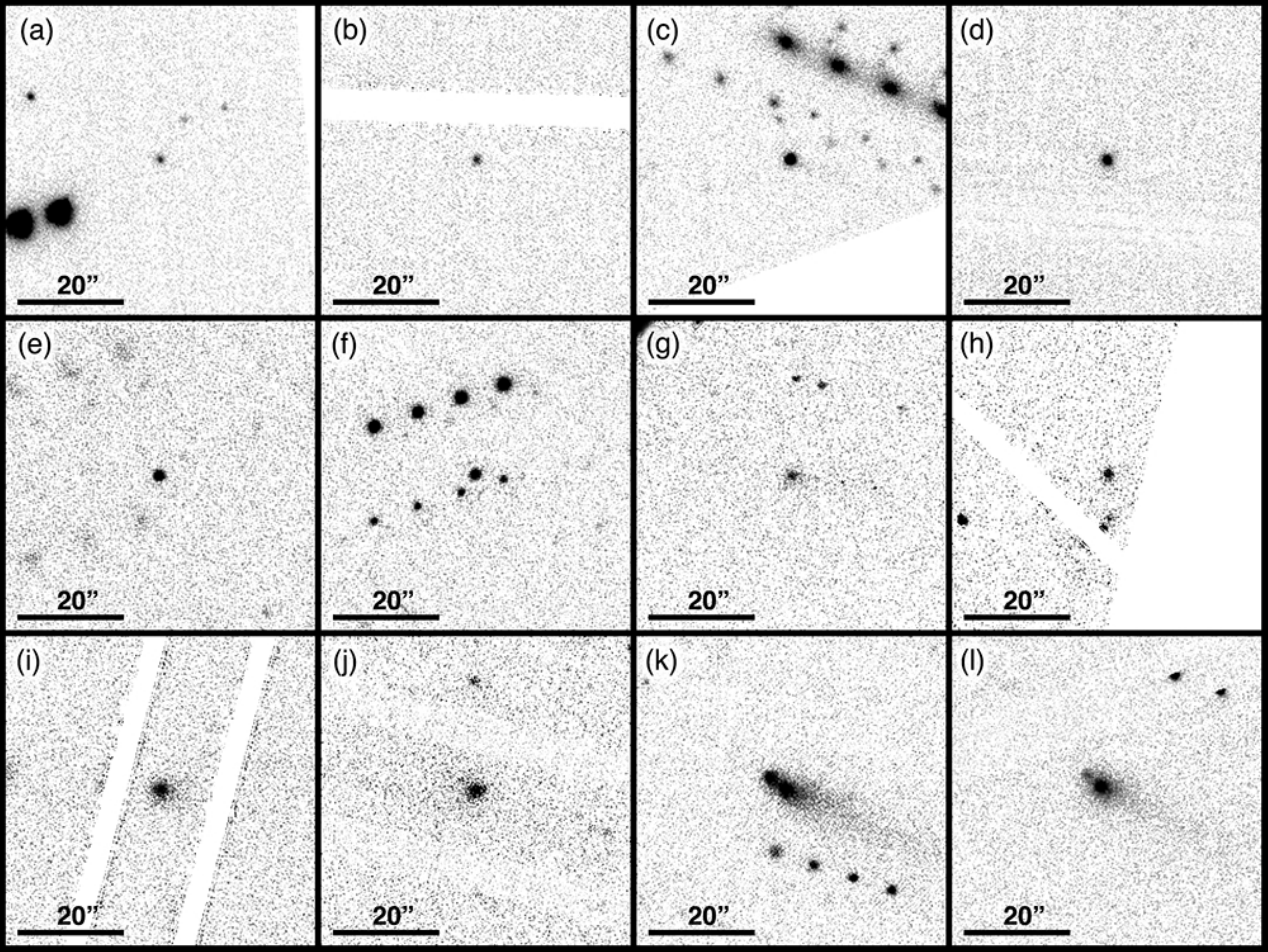}}
\caption{\small Stacked composite images of MBCs (center of each panel) observed by PS1 during the time period considered here.  Observations shown are of 133P/Elst-Pizarro on (a) 2012 May 20, (b) 2012 June 7, (c) 2013 August 31, and (d) 2013 September 27; 176P/LINEAR on (e) 2013 February 9, and (f) 2013 March 4; P/2012 T1 (PANSTARRS) on (g) 2012 October 6, (h) 2012 October 8, (i) 2012 October 20, and (j) 2012 December 7; and P/2013 R3 (Catalina-PANSTARRS) on (k) 2013 September 15, and (l) 2013 October 4.  Each panel is $60''\times60''$ in angular size with North at the top and East to the left.  Observational circumstances are listed in Table~\ref{table:ps1mbcobs}.  White strips visible in some panels are chip gaps or edges.
}
\label{figure:ps1mbcs}
\end{figure}

\begin{figure}
\centerline{\includegraphics[width=5in]{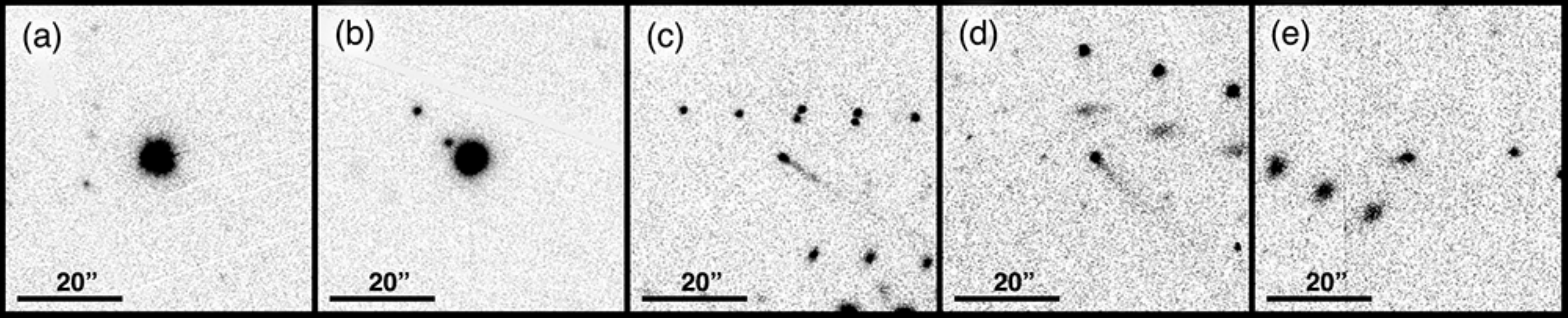}}
\caption{\small Stacked composite images of disrupted asteroids (center of each panel) observed by PS1 during the time period considered here.  Observations shown are of (596) Scheila on (a) 2013 February 9, and (b) 2013 March 4, and P/2013 P5 (PANSTARRS) on (c) 2013 August 15, (d) 2013 September 3, and (e) 2013 September 27.  Each panel is $60''\times60''$ in angular size with North at the top and East to the left.  Observational circumstances are listed in Table~\ref{table:ps1mbcobs}.  Observations of (596) Scheila are included for completeness, but as these detections did not meet our $S/N$ cut (Section~\ref{observations}), they are not included in any of the statistical analyses described in this work
}
\label{figure:ps1das}
\end{figure}

The vast majority of comets discovered by PS1 are Jupiter-family comets (JFCs) or long-period comets (LPCs) (Table~\ref{table:ps1discoveries}), but PS1 has also discovered two active Centaurs --- C/2011 P2 (PANSTARRS), and C/2013 P4 (PANSTARRS) --- and four active asteroids to date: 288P \citep[also known as asteroid (300163) 2006 VW$_{139}$;][]{hsi11c,hsi12b}, P/2012 T1 (PANSTARRS) \citep[][]{p2012t1}, P/2013 P5 (PANSTARRS) \citep[][]{p2013p5}, and P/2013 R3 (PANSTARRS) \citep[][]{p2013r3} (Table~\ref{table:knownaas}; Figures~\ref{figure:ps1mbcs} and \ref{figure:ps1das}).  Of these, 288P, P/2012 T1, and P/2013 R3 are likely to be ice-bearing MBCs \citep{hsi12b,hsi13a,jew14a}, and are all found in the outer main belt (Figure~\ref{figure:mbcs_aei}), with semimajor axes all beyond 3 AU, while P/2013 P5 is believed to be a disrupted asteroid \citep{jew13c,hai14,mor14}, and orbits near the innermost edge of the main belt.

In addition to PS1 discoveries, our data set also includes observations of known active asteroids, including 133P, 176P, and (596) Scheila (Table~\ref{table:ps1mbcobs}; Figures~\ref{figure:ps1mbcs} and \ref{figure:ps1das}).  None of these objects were flagged by MOPS as being potentially cometary in these PS1 observations, even though on one night, 133P did actually exhibit a marginally visible dust tail in a stacked composite image of all of the PS1 observations taken that night.

\begin{table}[ht]
\caption{PS1 Active Asteroid Observations (2012 May 20 - 2013 Nov 9)}
\smallskip
\scriptsize
\begin{tabular}{lclcrrrrccc}
\hline\hline
 \multicolumn{1}{c}{Object} &
 \multicolumn{1}{c}{Type$^a$} &
 \multicolumn{1}{c}{Obs.\ Date} &
 \multicolumn{1}{c}{Filt.} &
 \multicolumn{1}{c}{$m_V^b$} &
 \multicolumn{1}{c}{$R^c$} &
 \multicolumn{1}{c}{$\nu^d$} &
 \multicolumn{1}{c}{$S/N^e$} &
 \multicolumn{1}{c}{\tt psfq.$^f$} &
 \multicolumn{1}{c}{\tt psfex.$^g$} &
 \multicolumn{1}{c}{Active?$^h$} \\
 \hline
133P/Elst-Pizarro                       & MBC & 2012 May 20           & $g_{\rm P1}$           & 20.0 & 2.860 & 298.2 &   9.2 & 0.482 & 1.86 & no  \\ 
133P/Elst-Pizarro                       & MBC & 2012 Jun  7           & $r_{\rm P1}$           & 20.1 & 2.835 & 302.0 &   9.9 & 0.400 & 1.61 & no  \\ 
133P/Elst-Pizarro                       & MBC & 2013 Aug 31           & $w_{\rm P1}$           & 19.4 & 2.780 &  48.5 &  35.8 & 0.708 & 2.53 & yes \\ 
133P/Elst-Pizarro                       & MBC & 2013 Sep 27           & $w_{\rm P1}$           & 19.9 & 2.814 &  54.5 &  19.1 & 0.583 & 2.72 & no  \\ 
176P/LINEAR                             & MBC & 2013 Feb  9           & $w_{\rm P1}$           & 20.1 & 3.428 & 122.4 &  23.9 & 0.601 & 1.76 & no  \\ 
176P/LINEAR                             & MBC & 2013 Mar  4           & $w_{\rm P1}$           & 20.2 & 3.464 & 125.7 &  22.9 & 0.621 & 1.77 & no  \\ 
P/2012 T1 (PANSTARRS)                   & MBC & 2012 Oct  6$^\dagger$ & $r_{\rm P1}i_{\rm P1}$ & 20.6 & 2.415 &   7.5 &  20.6 & 0.527 & 3.00 & yes \\ 
P/2012 T1 (PANSTARRS)                   & MBC & 2012 Oct  8           & $r_{\rm P1}i_{\rm P1}$ & 20.8 & 2.415 &   8.1 &   7.1 & 0.514 & 2.78 & yes \\ 
P/2012 T1 (PANSTARRS)                   & MBC & 2012 Oct 20           & $w_{\rm P1}$           & 20.4 & 2.420 &  11.5 &  12.3 & 0.640 & 2.77 & yes \\ 
P/2012 T1 (PANSTARRS)                   & MBC & 2012 Dec  7           & $w_{\rm P1}$           & 21.0 & 2.456 &  25.3 &  10.5 & 0.679 & 3.15 & yes \\ 
P/2013 R3 (Catalina-PANSTARRS)          & MBC & 2013 Sep 15$^\dagger$ & $w_{\rm P1}$           & 20.5 & 2.218 &  14.0 &   9.2 & 0.749 & 4.38 & yes \\ 
P/2013 R3 (Catalina-PANSTARRS)          & MBC & 2013 Oct 04           & $g_{\rm P1}$           & 19.1 & 2.233 &  20.3 &  25.6 & 0.945 & 4.69 & yes \\ 
(596) Scheila$^\ddagger$                & DA  & 2013 Nov 19           & $r_{\rm P1}i_{\rm P1}$ & 15.6 & 3.144 & 124.9 & 269.0 & 0.975 & 2.83 & no  \\ 
(596) Scheila$^\ddagger$                & DA  & 2013 Dec 23           & $r_{\rm P1}$           & 15.0 & 3.189 & 130.6 & 204.6 & 0.986 & 2.42 & no  \\ 
P/2013 P5 (PANSTARRS)                   & DA  & 2013 Aug 15$^\dagger$ & $w_{\rm P1}$           & 21.0 & 2.147 & 272.8 &  21.0 & 0.607 & 3.16 & yes \\ 
P/2013 P5 (PANSTARRS)                   & DA  & 2013 Sep  3           & $w_{\rm P1}$           & 20.4 & 2.122 & 278.8 &  22.7 & 0.687 & 2.85 & yes \\ 
P/2013 P5 (PANSTARRS)                   & DA  & 2013 Sep 27           & $w_{\rm P1}$           & 20.6 & 2.091 & 286.6 &  11.2 & 0.540 & 3.20 & yes \\ 
\hline
\hline
\end{tabular}
\newline {$^a$ Type of active asteroid (MBC: main-belt comet; DA: disrupted asteroid)}
\newline {$^b$ Equivalent apparent $V$-band magnitude}
\newline {$^c$ Heliocentric distance at time of observation, in AU}
\newline {$^d$ True anomaly at time of observation, in degrees}
\newline {$^e$ Median signal-to-noise ratio of observation}
\newline {$^f$ {\tt psfquality} parameter value (as described in text)}
\newline {$^g$ {\tt psfextent} parameter value (as described in text)}
\newline {$^h$ Is activity visibly present in PS1 data?}
\newline {$^\dagger$ Date of discovery by PS1}
\newline {$^\ddagger$ Observations of (596) Scheila are included for completeness, but as these detections did not meet our $S/N$ cut (Section~\ref{observations}), they are not included in any of the statistical analyses described in this work.}
\label{table:ps1mbcobs}
\end{table}

PS1 has been exceptionally successful at discovering comets in general, and at discovering MBCs and disrupted asteroids in particular.  Of the eight currently recognized MBCs for which visible cometary activity has been observed (Ceres's activity was discovered spectroscopically), PS1 is credited with discovering the cometary activity of the last three to be identified, and is the only survey to date to have discovered multiple MBCs.  It has also discovered one of the only four currently recognized disrupted asteroids.  This discovery total is even more striking considering that PS1 has only discovered 56 comets in total as of 2014 April 15, giving it a ratio of active asteroid discoveries to total comet discoveries of $\sim$7\%, far larger than the overall ratio of known active asteroids to total comets ever discovered.  We believe that this high rate of active asteroid discoveries could be due to an overabundance of active asteroids among weakly active comets, and PS1's ability to detect weaker activity than most other surveys.  As more active asteroids are discovered, the validity of this hypothesis will become clearer.  In the meantime, however, we expect that future deep wide-field surveys that should be sensitive to even fainter and weaker comets, such as the currently ongoing PS1+PS2 survey, Subaru HyperSuprimeCam (HSC) survey, and Dark Energy Survey (DES), and the upcoming LSST survey, should have the opportunity to discover many more active asteroids, and should therefore ensure that mechanisms are in place for them to do so (cf.\ \ref{appendix:futurelessons}).

Sensitivity to weak cometary activity should yield other side benefits as well in the form of the detection of other interesting cometary cases.  For example, among its 56 total comet discoveries, PS1 has also discovered 2 active Centaurs, adding to a small but growing number of objects known to exhibit cometary activity in the outer solar system between Jupiter and Neptune. Studying these objects can provide insights into the transition of trans-Neptunian objects into JFCs, as well as into the nature of cometary activity at large heliocentric distances \citep[e.g.,][]{jew09,lin14}.  Surveys able to detect very faint and weak comets will also be sensitive to approaching comets that will eventually become extremely bright at large distances while they are still faint, such as C/2011 L4 (PANSTARRS), which PS1 discovered on 2011 June 6 when it was 7.9~AU from the Sun \citep{wai11}, and C/2012 S1 (ISON), which appeared cometary in PS1 data as early as 2012 January 28 when it was 8.4~AU from the Sun, but was not flagged by our comet detection algorithms at the time and not officially discovered until 2012 September 21 when it was 6.3~AU from the Sun \citep{nev12}.  Very early discoveries of comets like these enable their activity to be tracked over extremely long time periods, providing additional opportunities to study activity over a wide range of heliocentric distances, and improving our understanding of the composition of dynamically new comets \citep[e.g.,][]{mee13,oro13,li13,yan14}.

\section{DISCUSSION}
\label{section:discussion}

\subsection{Properties of the Known MBC Population}
\label{section:mbcpopulation1}

One of the primary goals of this work is to use PS1 data to place constraints on the size and distribution of the MBC population.  To do this, we first note that almost all of the currently known MBCs, and both of the MBCs discovered by PS1 during our period of interest, are found in the outer main asteroid belt, defined here as the region between the 5:2 and 2:1 mean-motion resonances with Jupiter at 2.824~AU and 3.277~AU, respectively.  Thus, for this analysis, this is where we will focus our attention.
The two MBCs that are found in the middle main belt (between the 3:1 and 5:2 mean-motion resonances with Jupiter at 2.501~AU and 2.824~AU, respectively), 259P/Garradd and Ceres, are therefore excluded from much of the discussion that follows.  We note however that both are unusual compared to other MBCs for other additional reasons, such as the facts that 259P is suspected of being an interloper from elsewhere in the asteroid belt or even elsewhere in the solar system, and that Ceres's large size likely means it has very different physical conditions compared to the much smaller MBCs \citep[e.g.][]{mcc05,mcc11,cas11}.  As such, it is unlikely that any attempt to draw general conclusions about the overall MBC population based on these two objects would be particularly meaningful at this time.

\begin{figure}
\centerline{\includegraphics[width=6.0in]{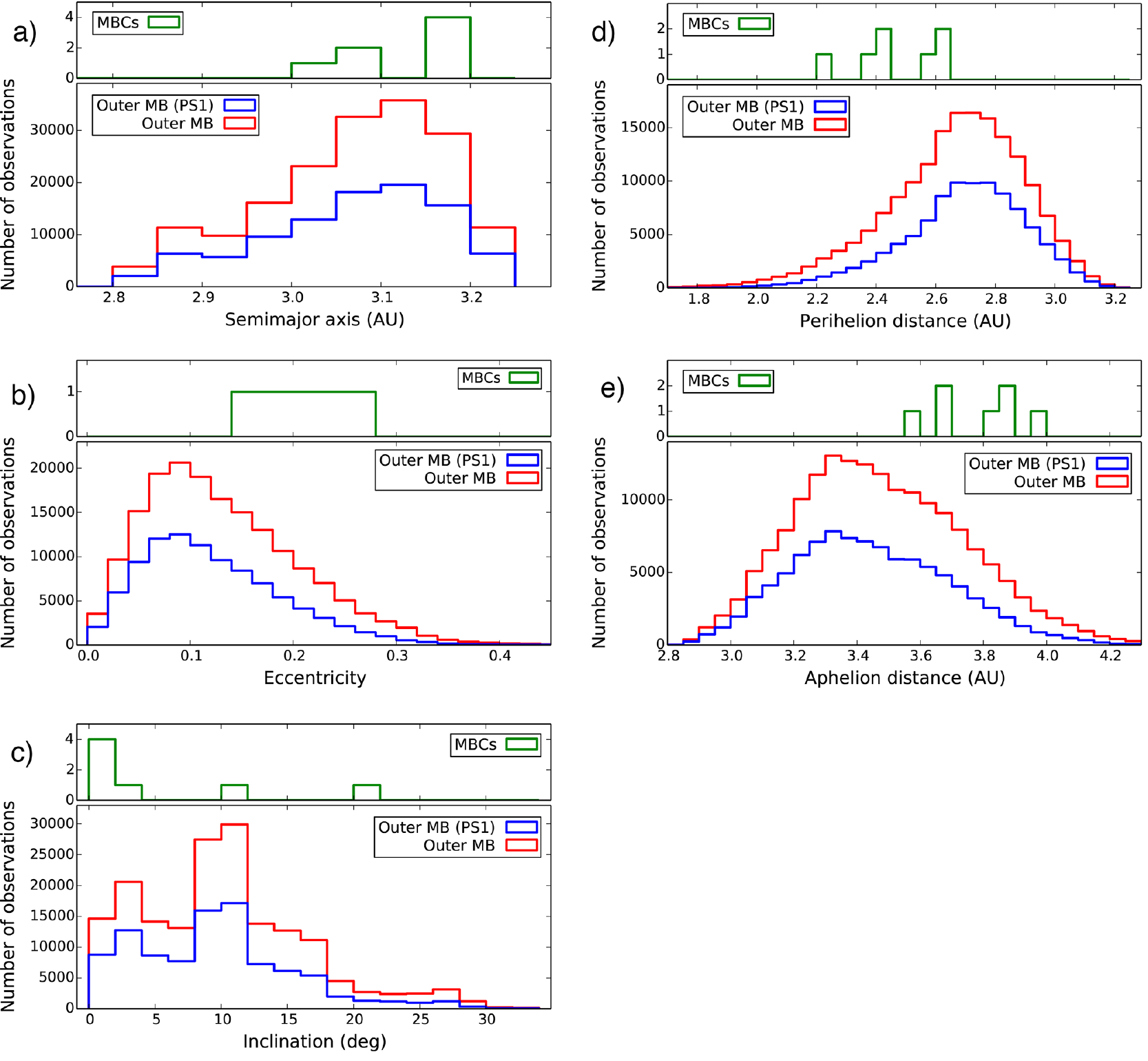}}
\caption{\small Histograms showing the (a) semimajor axis, (b) eccentricity, (c) inclination, (d) perihelion distance, and (e) aphelion distances of the total known outer main-belt asteroid population (red lines), the outer main-belt asteroid population observed by PS1 (blue lines), and the known MBCs (green lines).
  }
\label{figure:hist_outer_aeiqQ}
\end{figure}

Considering only the outer main belt, we replot the orbital element distributions of the known asteroid population and the population observed by PS1 previously plotted in Figure~\ref{figure:hist_aeiH} (Figure~\ref{figure:hist_outer_aeiqQ}) along with the orbital element distributions of the known outer-belt MBC population.  While the semimajor axis distribution of the known MBC population is similar to that of the underlying outer main-belt population (Figure~\ref{figure:hist_outer_aeiqQ}a), the MBC population is notably skewed towards higher eccentricities and lower inclinations (Figures~\ref{figure:hist_outer_aeiqQ}b,c).  

Two-sample Kolmogorov-Smirnov tests confirm these qualitative assessments, indicating that there is a $\sim$30\% probability that the semimajor axis distances of the known MBCs are drawn from the same distribution as those of the outer main-belt asteroid population, while there are just $\sim$0.2\% and $\sim$2\% probabilities that the eccentricities and inclinations of the MBCs were drawn from the same distribution as those of the outer main-belt asteroid population.  The distribution of the currently known MBCs could be skewed due to the association of at least three of the known MBCs (133P, 176P, and 288P) with the moderate-eccentricity, low-inclination Themis family \citep{hsi04,hsi11a,hsi12b}, but there may also be physical reasons for the eccentricity and inclination distribution of the known MBCs.

\citet{far92} found moderately elevated collision rates for asteroids at low inclinations, as well as for members of asteroid families such as the Themis family.  If small impacts are needed to trigger activity \citep[transforming a previously inactive asteroid with subsurface ice into an actively sublimating MBC; cf.][]{hsi04}, it might then be reasonable to expect an excess of MBCs at low inclinations \citep{hsi09}.  The relatively large inclinations of disrupted asteroids believed to have undergone collisions ($i=14.7^{\circ}$ for Scheila, and $i=9.7^{\circ}$ for P/2012 F5) seem to be at odds with this hypothesis, but given that these are the only two asteroids likely to have experienced impacts that have been observed to date, we consider attempts to interpret their inclination distribution at this time to be premature.

On the other hand, the skewing of the outer main-belt MBC population towards larger eccentricities relative to the underlying asteroid population has not been previously noted.  This trend could indicate that the MBCs may in fact be drawn from a different dynamical population than the underlying asteroids, namely a population of highly evolved JFCs which happen to retain slightly larger eccentricities as remnants from their former cometary orbits.  Dynamical integrations (using gravitational forces only) by \citet{jfer02}, which were able to produce an orbit with a 133P-like semimajor axis and eccentricity from JFC 503D/Pigott, initially appear to support this possibility.  However, the inability of those simulations to reproduce 133P's low inclination (instead producing large inclinations of $\sim25^{\circ}-30^{\circ}$) is inconsistent with the observed excess of MBCs at small inclinations.  Alternatively, this eccentricity distribution could be due to an observational selection effect since outer main-belt asteroids with larger eccentricities will reach smaller heliocentric distances, where activity may be easier to detect, than the general population.  However, while the average heliocentric distance of the outer main-belt MBCs when they were discovered to be cometary was ${\bar R_{disc}}\sim2.5$~AU, and all were discovered to be cometary at heliocentric distances of $R_{disc}<2.7$~AU (Table~\ref{table:mbcdiscoveries}), dust emission events from both disrupted asteroids in the outer main belt (Scheila and P/2012 F5) were discovered at $R_{disc}\sim3.0$~AU \citep{lar10,gib12}, where both Scheila ($e=0.165$) and P/2012 F5 ($e=0.042$) have eccentricities lower than that of most outer main-belt MBCs.  These results suggest that observational selection is unlikely to be responsible for the apparent prevalence of MBC activity at preferentially small heliocentric distances, or equivalently, the prevalence of high-eccentricity outer main-belt MBCs.

Instead, we consider the possibility that the observed eccentricity distribution of the MBCs could be related to energy considerations.  When we examine the distributions of perihelion and aphelion distances for outer main-belt asteroids compared to those of the MBCs (Figure~\ref{figure:hist_outer_aeiqQ}d,e), we see that the modestly larger eccentricities of the MBCs cause them to have preferentially smaller perihelion distances and larger aphelion distances.  This means that they reach higher maximum temperatures and lower minimum temperatures than other asteroids with the same semimajor axis distances but smaller eccentricities.  We examine the thermal implications of this observation below (Section~\ref{section:thermal}).

\subsection{Inferred Properties of the Total MBC Population}
\label{section:mbcpopulation2}

\begin{figure}
\centerline{\includegraphics[width=4.0in]{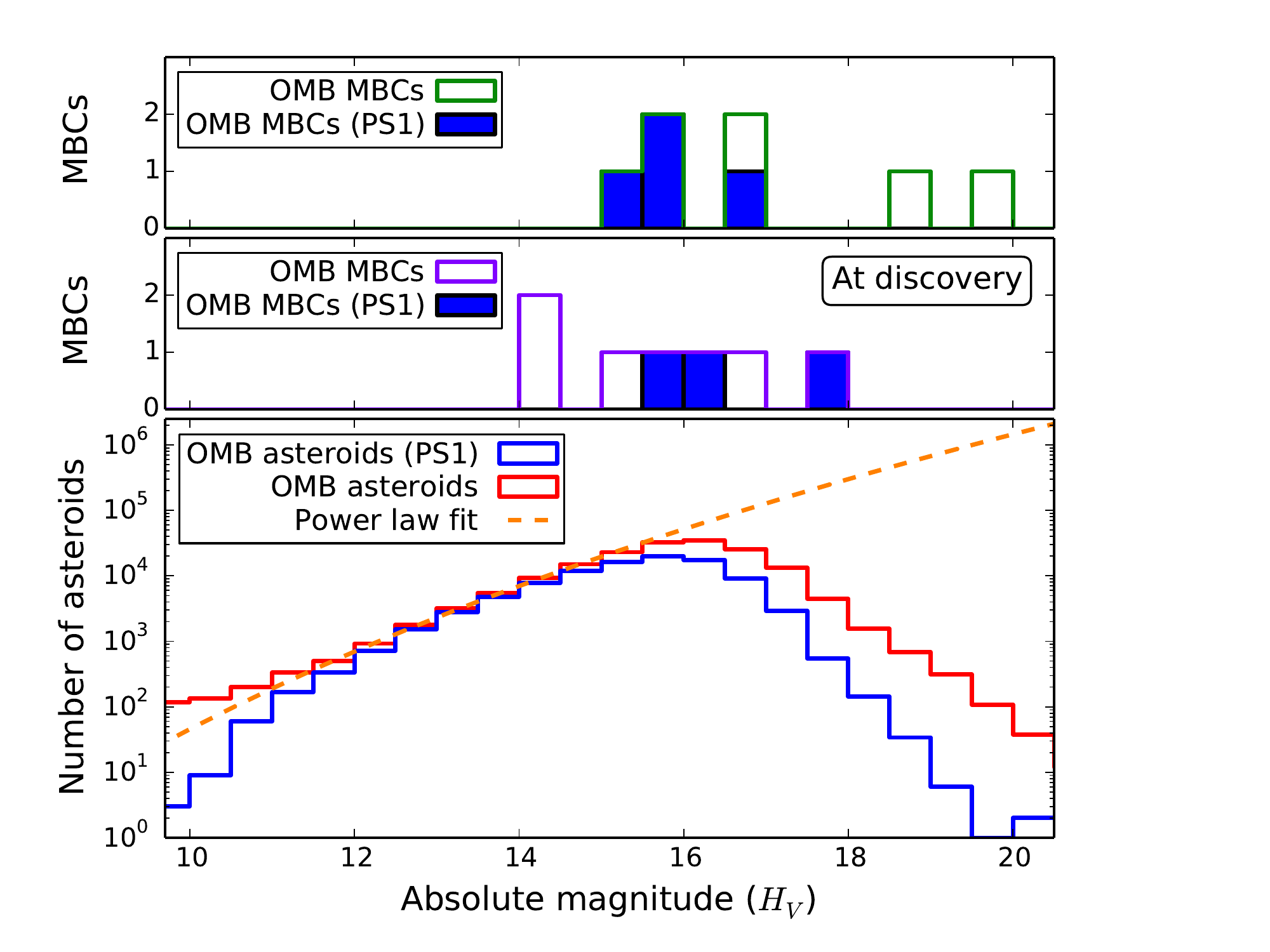}}
\caption{\small Histogram of the $V$-band absolute magnitudes of all known asteroids (red line) and all asteroids observed by PS1 (blue line) in the outer main belt ($2.824~{\rm AU} < a < 3.277~{\rm AU}$).  We also plot a power law fit (orange dashed line) to the known population from $H_V=12$~mag to $H_V=15$~mag, and the distribution of the PS1-observed MBCs (filled blue bars), the $H_V$ magnitudes of the inactive nuclei of the known MBCs (green line), and the effective $H_V$ magnitudes of the known MBCs at the times of their discoveries when they were active (purple lines).
  }
\label{figure:hist_outer_hmag}
\end{figure}

Figure~\ref{figure:hist_outer_hmag} shows the distribution of absolute magnitudes of all known asteroids and all unique asteroids observed by PS1 in the outer main belt.  In addition, we plot a power law fit to the known asteroid population, fitting over the absolute magnitude range of $12<H_V<15$ (over which the known main-belt population appears to be complete), finding a power law index of $-2.4$, consistent with the results of \citet{dem13}.  We find that the PS1-observed population appears incomplete relative to the total known population for asteroids with absolute magnitudes $H_V>14.5$, while both the PS1-observed population and total known population appear incomplete relative to the power law fit for asteroids with $H_V>15.0$ (although the PS1-observed population is more incomplete for asteroids with larger $H_V$ magnitudes).

Figure~\ref{figure:hist_outer_hmag} also shows the distribution of $H_V$ magnitudes of the nuclei of the known MBCs in the outer main belt (where some are lower limits) as well as the nuclei of the PS1-observed population of MBCs in the outer main belt.  The absolute magnitudes of the nuclei of the MBCs in this region range from $H_V=15.5$~mag down to $H_V=19.5$~mag, and so we extrapolate the power law fit and integrate the total number of objects to estimate a total population size of $\sim2.4\times10^6$ objects with $12<H_V<19.5$~mag.  Using an apparent $V$-band solar magnitude of $m_{\odot}=-26.70$~mag \citep{har80}, we compute equivalent effective diameters (in km), $d_e$, for the nuclei of the known MBCs using
\begin{equation}
p_Vd_e^2 = (8.96\times10^{16})\times10^{0.4(m_{\odot}-H_V)}
\end{equation}
where we assume a typical MBC nucleus albedo of $p_V=0.05$ \citep[][]{hsi09b,bau12}, finding an equivalent diameter range of $d_e\sim5$~km down to $d_e\sim0.8$~km.

\setlength{\tabcolsep}{3pt}
\begin{table}[ht]
\caption{Main-Belt Comet Discovery Circumstances}
\smallskip
\scriptsize
\begin{tabular}{llcrcrrcccc}
\hline\hline
 \multicolumn{1}{c}{Object Name} &
 \multicolumn{1}{c}{Disc.\ Date$^a$} &
 \multicolumn{1}{c}{Tel.$^b$} &
 \multicolumn{1}{c}{$H_V^c$} &
 \multicolumn{1}{c}{$R^d$} &
 \multicolumn{1}{c}{$\Delta^e$} &
 \multicolumn{1}{c}{$\alpha^f$} &
 \multicolumn{1}{c}{$m_{V,Disc}^g$} &
 \multicolumn{1}{c}{$H_{V,Disc}^h$} &
 \multicolumn{1}{c}{$\Delta m_{V,Disc}^i$} &
 \multicolumn{1}{c}{Refs.$^j$} \\ 
\hline
(1) Ceres                       & 2011 Nov 23$^\dagger$           & {\it Herschel} &     3.3 & 2.943 & 2.510 & 18.8 &  --- &  --- &  ~~--- & [1]  \\
133P/Elst-Pizarro = (7968)      & 1996 Jul 14$^\dagger$           & ESO 1.0m       &    15.9 & 2.654 & 1.762 & 13.1 & 18.3 & 14.2 &  ~~1.7 & [2]  \\
176P/LINEAR = (118401)          & 2005 Nov 26$^\dagger$           & Gemini-N       &    15.5 & 2.588 & 1.817 & 16.4 & 19.5 & 15.3 &  ~~0.2 & [3]  \\
238P/Read                       & 2005 Oct 24                     & SW 0.9m        &    19.4 & 2.416 & 1.463 &  8.7 & 20.2 & 16.9 &  ~~2.5 & [4]  \\
259P/Garradd                    & 2008 Sep  2                     & SS 0.5m        &    20.1 & 1.817 & 0.938 & 21.9 & 18.5 & 16.3 &  ~~3.8 & [5]  \\
288P/2006 VW$_{139}$ = (300163) & 2011 Nov  5$^{\dagger\ddagger}$ & PS1            &    16.9 & 2.496 & 1.517 &  4.6 & 18.8 & 15.5 &  ~~1.4 & [6]  \\
P/2010 R2 (La Sagra)            & 2010 Sep 14                     & LS 0.45m       &    18.8 & 2.644 & 1.743 & 12.0 & 18.4 & 14.4 &  ~~4.4 & [7]  \\
P/2012 T1 (PANSTARRS)           & 2012 Oct  6$^\ddagger$          & PS1            & $>$16.9 & 2.415 & 1.540 & 14.4 & 20.0 & 16.3 & $>$0.6 & [8]  \\
P/2013 R3 (Catalina-PANSTARRS)  & 2013 Sep 15$^\ddagger$          & PS1            & $>$17.5 & 2.218 & 1.254 &  9.8 & 20.4 & 17.5 & $>$0.0 & [9]  \\ 
\hline
\hline
\end{tabular}
\newline {$^a$ UT date of discovery of object in cases of objects discovered as comets, or UT date of discovery of cometary activity in cases of objects previously known as asteroidal objects.}
\newline {$^b$ Telescope used to discover cometary activity ({\it Herschel}: {\it Herschel Space Observatory}; ESO 1.0m: European Southern Observatory 1.0m (La Silla, Chile), Gemini-N:  Gemini North, Gemini Observatory (Mauna Kea, Hawaii); SW 0.9m: Spacewatch 0.9m (Kitt Peak, Arizona); SS 0.5m: 0.5m Uppsala Southern Schmidt Telescope (Siding Spring, Australia); PS1: Pan-STARRS1 (Haleakala, Hawaii); LS 0.45m: La Sagra Observatory 0.45m telescope (Spain)}
\newline {$^c$ Absolute magnitude of nucleus}
\newline {$^d$ Heliocentric distance, in AU, of object at time of discovery of cometary activity}
\newline {$^e$ Geocentric distance, in AU, of object at time of discovery of cometary activity}
\newline {$^f$ Solar phase angle, in degrees, of object at time of discovery of cometary activity}
\newline {$^g$ Equivalent apparent $V$-band magnitude of object reported at time of discovery of cometary activity}
\newline {$^h$ Equivalent absolute $V$-band magnitude of object at time of discovery of cometary activity, assuming G=0.15}
\newline {$^i$ Observed photometric excess of object at time of discovery of cometary activity relative to expected brightness of inactive nucleus, in mag}
\newline {$^j$ References:
[1] \citet{ted04,kup14}; 
[2] \citet{els96,hsi10b};
[3] \citet{hsi09b,hsi11a};
[4] \citet{rea05,hsi11b};
[5] \citet{gar08,mac12};
[6] \citet{hsi11c,hsi12b}; Hsieh et al., in prep;
[7] \citet{nom10,hsi12c,hsi14b};
[8] \citet{p2012t1,hsi13a};
[9] \citet{p2013r3,jew14a}
}
\newline {$^\dagger$ Previously known as an inactive asteroid prior to discovery of comet-like activity}
\newline {$^\ddagger$ Cometary activity discovered by PS1}
\label{table:mbcdiscoveries}
\end{table}

Following the Bayesian statistical analysis of \citet{was13}, we assume a prior probability distribution on the fraction, $f$, of active MBCs in the outer main belt with activity levels detectable by PS1 \citep[i.e., equivalent to mass loss rates on the order of $\sim0.1-1$~kg~s$^{-1}$; e.g.,][]{hsi09a,hsi11a,mor11a,mor13,lic13a,jew14b}, of
\begin{equation}
P(f) = -{1\over f\log f_{\rm min}}
\end{equation}
where $f_{\rm min}>0$ is the minimum assumed value of $f$ (allowed to be arbitrarily small), and $f_{\rm min}<f<1$, and approximating the likelihood probability distribution function for a general sample, $S$, as a Poisson distribution, given a very large number, $N$, of asteroids, we adopt
\begin{equation}
P(f|S) = {P(S|f)P(f)\over \int_{f_{\rm min}}^1 \! P(S|f)P(f)\,\mathrm{d}f}\propto f^{n-1}\exp(-NCf)
\end{equation}
as the posterior probability distribution on $f$ given our results, where $n=2$ is the number of active MBCs detected in our sample, and $C$ is the completeness or efficiency of our MBC-detection procedures.  In Section~\ref{section:efficiency}, we calculated an efficiency rate of $C=0.7$ for our comet-detection procedures, but noted that this was only an upper limit.  For the purposes of this analysis, we conservatively assume $C=0.5$.  We can then compute the expected fraction, $f_{50}$, of MBCs in the outer main belt by numerically solving
\begin{equation}
\int_{f_{\rm min}}^{f_{50}}\! P(f|S)\,df = 0.50 ~,
\label{equation:f50}
\end{equation}
as well as the 95\% confidence upper limit, $f_{95}$, by solving
\begin{equation}
\int_{f_{\rm min}}^{f_{95}}\! P(f|S)\,df = 0.95 ~.
\label{equation:f95}
\end{equation}
Our data sample contains two active MBC detections (P/2012 T1 and P/2013 R3) in a total sample of 96\,162 outer main-belt asteroids with $12<H_V<19.5$, giving us an expected fraction of $f_{50}=33$~MBCs for every $10^6$ outer main-belt asteroids, and a 95\% confidence upper limit fraction of $f_{95}=82$~MBCs for every $10^6$ outer main-belt asteroids.  Given a total outer main-belt population size of $2.4\times10^6$ objects with $12<H_V<19.5$, these results correspond to an expected total of $\sim$80~MBCs and an upper limit total of $\sim$200~MBCs within our specified absolute magnitude limits, assuming $C=0.5$.

\begin{table}[ht]
\caption{Ranges of Activity for Known Main-Belt Comets}
\smallskip
\scriptsize
\begin{tabular}{lccccc}
\hline\hline
 \multicolumn{1}{c}{Name} &
 \multicolumn{1}{c}{$R$ range$^a$} &
 \multicolumn{1}{c}{$\nu$ range$^b$} &
 \multicolumn{1}{c}{$f_{actv}^c$} &
 \multicolumn{1}{c}{Refs.$^d$} \\ 
\hline
(1) Ceres                       & $2.62-2.72$ & $279.3-312.6$ & 0.09 & [1]  \\ 
133P/Elst-Pizarro = (7968)      & $2.64-3.25$ & $349.9-109.0$ & 0.26 & [2]  \\ 
176P/LINEAR = (118401)          & $2.59-2.60$ & $10.1-18.6$   & 0.02 & [3]  \\ 
238P/Read                       & $2.42-2.57$ & $306.1-43.9$  & 0.17 & [4]  \\ 
259P/Garradd                    & $1.82-1.97$ & $18.5-48.6$   & 0.04 & [5]  \\ 
288P/2006 VW$_{139}$ = (300163) & $2.45-2.58$ & $12.2-47.4$   & 0.07 & [6]  \\ 
P/2010 R2 (La Sagra)            & $2.63-2.77$ & $12.9-53.5$   & 0.09 & [7]  \\ 
P/2012 T1 (PANSTARRS)           & $2.41-2.53$ & $7.4-41.4$    & 0.06 & [8]  \\ 
P/2013 R3 (Catalina-PANSTARRS)  & $2.22-2.24$ & $14.0-22.9$   & 0.01 & [9]  \\ 
\hline
\hline
\end{tabular}
\newline {$^a$ Heliocentric distance range, in AU, over which activity has been observed}
\newline {$^b$ True anomaly range, in deg, over which activity has been observed}
\newline {$^c$ Time fraction of object's orbit during which activity has been observed}
\newline {$^d$ References:
[1] \citet{kup14}; 
[2] \citet{els96,hsi04,hsi10b,hsi13b,low05,kal11,jew14b};
[3] \citet{hsi06a,hsi11a,hsi14};
[4] \citet{rea05,hsi09a,hsi11b};
[5] \citet{gar08,jew09a,mac12};
[6] \citet{hsi11c,hsi12b};
[7] \citet{nom10,hsi12c};
[8] \citet{p2012t1,hsi13a};
[9] \citet{p2013r3,lic13b,jew14a};
[10] \citet{lar10,jew11b,mor11b};
[11] \citet{bir10,jew10,jew11a,jew13a,hai12};
[12] \citet{gib12,ste12b}; this work (Section~\ref{section:ps1blindspots});
[13] \citet{p2013p5,jew13c}
[14] \citet{mai10};
}
\label{table:activeranges}
\end{table}

We note, however, that MBC activity does not appear to be uniformly distributed around their orbits.  In Figure~\ref{figure:hist_rdmv}, we see that PS1 observations of MBC activity is confined to a narrow range of true anomalies between $\nu=0^{\circ}$ and $\nu=45^{\circ}$.  The range of true anomalies over which all reported observations of MBCs show them to be active is somewhat larger.  With the exception of Ceres, though, which we already have judged to be a special case, the range is still largely limited to the post-perihelion quadrant of each orbit ($\nu\sim0^{\circ}-90^{\circ}$), with occasional instances of pre-perihelion activity being observed (Table~\ref{table:activeranges}).  We note, however, that not all of the activity exhibited over this range is detectable by PS1.  For example, no activity was detected by PS1 for 133P on either 2013 August 31 or 2013 September 27 (although a very faint dust tail was visible in stacked data from 2013 August 31) (Table~\ref{table:ps1mbcobs}), despite both of these observations occurring during a portion of 133P's orbit during which it has been active during previous apparitions \citep{hsi04,hsi10b}.  With these considerations in mind, if we only focus on the range of true anomalies where we would expect MBCs to be active {\it and} where PS1 would be able to detect that activity (assumed here to be $\nu=0^{\circ}$ to $\nu=45^{\circ}$), we are left with a sample of 30\,653 objects with $12<H_V<19.5$~mag which were observed by PS1 at least once during the true anomaly range in question, of which two were observed to be active.  Solving Equations~\ref{equation:f50} and \ref{equation:f95} for this subsample of objects, we find an expected fraction of $f_{50}=59$~MBCs for every $10^6$ outer main-belt asteroids and an upper limit of $f_{95}=96$~MBCs for every $10^6$ outer main-belt asteroids \citep[compatible with the results of][]{was13}, assuming $C=0.5$, corresponding to a total expected population of $\sim140$~MBCs and an upper limit of $\sim230$~MBCs within our specified absolute magnitude limits.

Reality is likely even more complex though.  There are a number of other physical constraints, such as restrictions on inclination, family membership, albedo, and spectral type \citep[cf.][]{hsi09,was13}, that we could consider to further refine our MBC population estimates.  Realistically, however, we have a very poor understanding of how any of these factors are truly related to MBC abundance, and so to keep our analysis here as simple as possible, we do not consider any of those issues at this time.  The absolute magnitude distribution of the MBC population is also poorly understood, as the nucleus sizes of many of the currently known MBCs have yet to be measured.  Furthermore, many MBCs brighten significantly while they are active (Table~\ref{table:mbcdiscoveries}; Figure~\ref{figure:hist_outer_hmag}), sometimes giving them effective absolute magnitudes several magnitudes brighter than their absolute nuclear magnitudes.  The effect of this behavior on the statistical analyses of MBC surveys is unclear at this time, but should be considered more in the future as more absolute magnitudes of MBC nuclei are determined, enabling the distribution of the magnitudes and decay rates of their photometric enhancements when active to become better understood.

Finally, we note that while PS1 did not detect any MBCs in the inner or middle regions of the main asteroid belt, where we observed 115\,088 and 121\,206 asteroids, respectively, it does not mean that MBCs do not exist in these regions, as clearly demonstrated by the existence of 259P and Ceres.  The active behavior patterns and observability of any MBCs in these regions may differ from those in the outer main belt and so it is unclear whether only considering asteroids with $0^{\circ}<\nu<45^{\circ}$, as we did in our analysis of the outer main belt, is similarly justified for the inner and middle main belt regions.  As such, if we consider the entire population of asteroids we observed in each region and assume $C=0.50$ as before, we find 95\% confidence upper limits of $f_{95}=24$~MBCs per $10^6$ inner main-belt asteroids and $f_{95}=23$~MBCs per $10^6$ middle main-belt asteroids given a detection rate of $n=0$ MBCs for these regions.  For reference, following our analysis of the outer main belt and but instead considering the samples of 17\,570 inner main-belt asteroids and 19\,181 middle main-belt asteroids with $12<H_V<19.5$~mag that we observed at $0^{\circ}<\nu<45^{\circ}$, we find upper limits of $f_{95}=67$~MBCs per $10^6$ inner main-belt asteroids and $f_{95}=66$~MBCs per $10^6$ inner main-belt asteroids with $12<H_V<19.5$~mag.

\subsection{Thermal Considerations}
\label{section:thermal}

\subsubsection{Background}
\label{section:backgroundthermal}

Numerical thermal models have indicated that water ice can in fact survive on main-belt objects over Gyr timescales \citep[e.g.,][]{fan89,sch08,pri09}, suggesting that dynamical results indicating that MBCs likely formed where we see them today \citep[e.g.,][]{hag09} are physically compatible with the present-day ability of these objects to exhibit cometary activity.  Even if ice preserved in primordial bodies is now only found at large depths below their surfaces, the catastrophic disruption of such bodies to form asteroid families provides a mechanism for producing present-day small bodies with ice in shallow subsurface layers where it can be easily excavated by small impacts \citep[e.g.,][]{hsi09,nov12}.  \citet{cap12} further showed that dust emission comparable to that observed for MBCs could be produced by such excavations of subsurface water ice on a main-belt asteroid.

The mechanism for the modulation of observed MBC activity has remained somewhat mysterious, however.  Following the re-discovery of active dust emission by 133P in 2002, \citet{hsi04} proposed that activity could be modulated by seasonal variations in solar illumination of an isolated active site, similar to seasonal temperature variations on Earth, assuming non-zero obliquity.  In the case of 133P, estimates of its pole orientation support this hypothesis \citep{tot06,hsi10b}.  As more MBCs have been discovered, however, an unmistakable pattern has emerged. Nearly all of the known MBCs, except Ceres, have been observed to exhibit activity close to or shortly after perihelion (Table~\ref{table:activeranges}), which is inconsistent with the aforementioned seasonal modulation hypothesis which predicts that activity for a MBC should peak near the summer solstice of the hemisphere of the object where the isolated active site is located.

There is no reason to expect any correlation between an object's solstices and its position in its orbit, and we would therefore expect MBC activity to be randomly distributed in true anomaly space, including near aphelion, but this is not what is observed.  We do find a wide range of true anomalies for disrupted asteroids during their active periods, indicating that collisional and rotational disruption events are minimally dependent on the orbital position of the affected object, as expected.  Conversely, the strong correlation of volatile-driven activity on MBCs with their perihelion passages suggests that MBC activity is dependent on orbital position, and such a dependency points to activity modulation that is dependent on heliocentric distance, rather than an object's obliquity and pole orientation and the specific location of its active site.

\subsubsection{Sublimation}
\label{section:sublimation}

The energy balance for an inert grey-body at a given distance from the Sun is given by
\begin{equation}
{F_{\odot}\over R^2}(1-A) = \chi\varepsilon\sigma T_{eq}^4
\label{equation:greybody}
\end{equation}
from which the equilibrium surface temperature, $T_{eq}(R)$, of that body is determined.  Here, $F_{\odot}=1360$~W~m$^{-2}$ is the solar constant, $R$ is the heliocentric distance of the object in AU, $A=0.05$ is the assumed Bond albedo of the body, $\chi$ describes the distribution of solar heating over an object's surface ($\chi=1$ for a flat slab facing the Sun where this so-called subsolar approximation produces the maximum attainable temperature for an object, $\chi=\pi$ for the equator of a rapidly rotating body with zero axis tilt, and $\chi=4$ for an isothermal sphere, as in the limiting approximation of an extremely fast rotator and strong meridional heat flux), $\sigma$ is the Stefan-Boltzmann constant, and $\varepsilon=0.9$ is the assumed effective infrared emissivity.

Including water ice sublimation as an additional energy balance consideration gives 
\begin{equation}
{F_{\odot}\over R^2}(1-A) = \chi\left[{\varepsilon\sigma T^4 + L f_D\dot m_{w}(T)}\right]
\label{equation:sublim1}
\end{equation}
where $L=2.83$~MJ~kg$^{-1}$ is the latent heat of sublimation of water ice, which is nearly independent of temperature, $f_D$ describes the reduction in sublimation efficiency caused by the diffusion barrier presented by the growing rubble mantle (discussed below; Section~\ref{section:mantles}), where $f_D=1$ in the absence of a mantle, and $\dot m_w$ is the water mass loss rate due to sublimation of surface ice.
\cite{del71} suggest that the latent heat of sublimation for water ice has a small but notable temperature dependence, but their quoted values are inconsistent with the modern literature.  
The sublimation rate of ice into a vacuum is given by
\begin{equation}
\dot m_{w} = P_v(T) \sqrt{\mu\over2\pi k T}
\label{equation:sublim3}
\end{equation}
where $\mu=2.991\cdot 10^{-26}$~kg is the mass of one water molecule, and $k$ is the Boltzmann constant, and the equivalent ice recession rate, $\dot \ell_{i}$, corresponding to $\dot m_{w}$ is given by
$\dot \ell_{i} = \dot m_{w}/ \rho$,
where $\rho$ is the bulk density of the object.  Then, using the Clausius-Clapeyron relation
\begin{equation}
P_v(T) = 611 \times \exp\left[{{\Delta H_{subl}\over R_g}\left({{1\over 273.16} - {1\over T}}\right)}\right]
\label{equation:sublim4}
\end{equation}
to compute the vapor pressure of water, $P_v(T)$, in Pa, where $\Delta H_{subl}=51.06$~MJ~kmol$^{-1}$ is the heat of sublimation for ice to vapor and $R_g=8314$~J~kmol$^{-1}$~K$^{-1}$ is the ideal gas constant, we iteratively calculate the equilibrium temperature of a sublimating grey-body at a given heliocentric distance, assuming $f_D=1$.  We only consider water sublimation here because previous analyses have demonstrated that water ice is likely to be the only surviving volatile material in main-belt objects \citep[e.g.,][]{pri09}.  

\begin{figure}
\centerline{\includegraphics[width=3.5in]{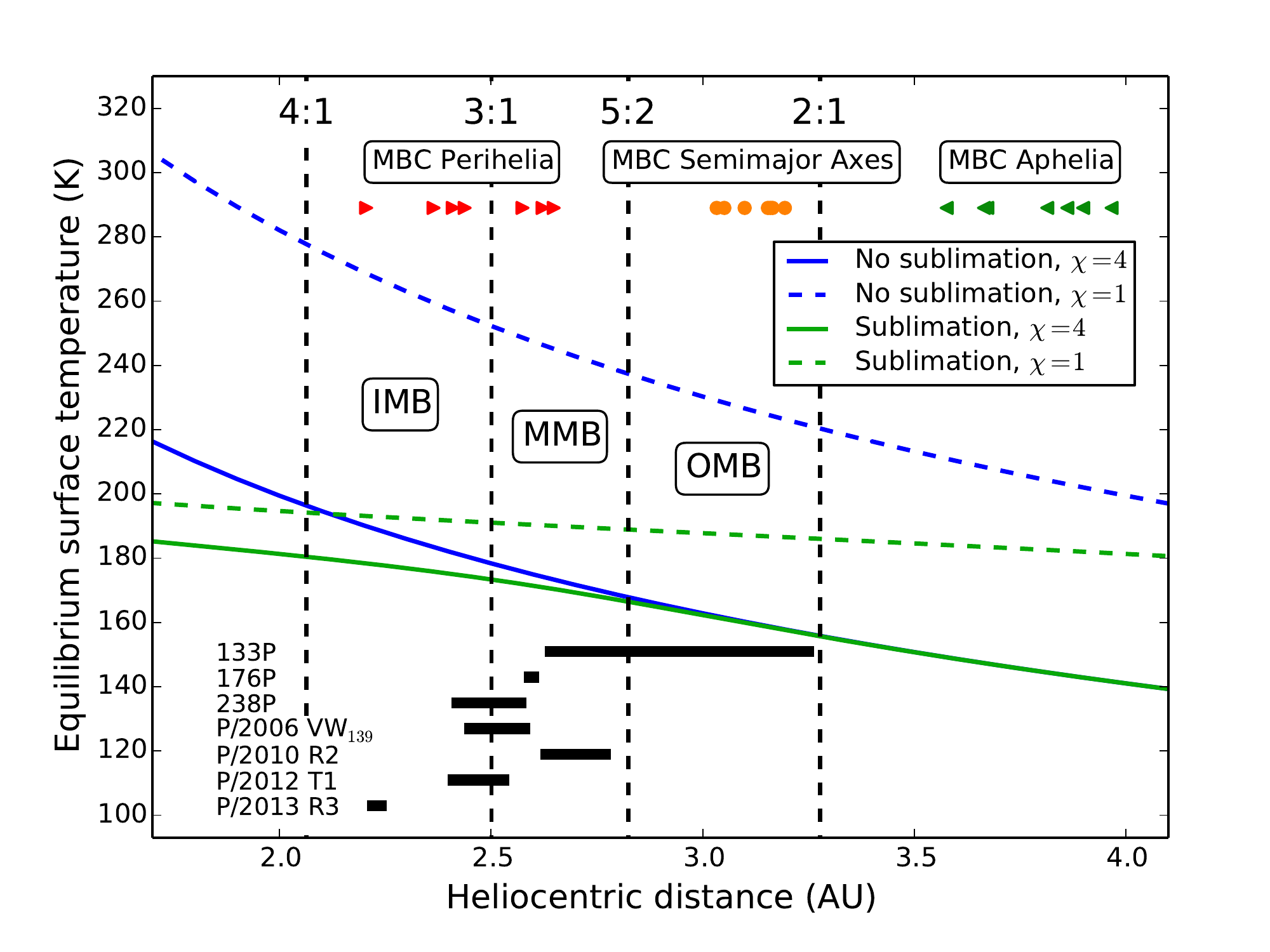}}
\caption{\small Equilibrium surface temperature of non-sublimating (blue lines) and sublimating (green lines) grey bodies as a function of heliocentric distance over the range of the main asteroid belt for water ice sublimation. Temperatures calculated using the isothermal approximation ($\chi=4$) and the subsolar approximation ($\chi=1$) are marked with solid and dashed lines, respectively, while the semimajor axis ranges of the inner, middle, and outer main belt are labeled IMB, MMB, and OMB, respectively.  The positions of the major mean-motion resonances with Jupiter (4:1, 3:1, 5:2, and 2:1) that delineate the various regions of the main asteroid belt are shown with vertical dashed black lines.  Also plotted are the perihelion distances (red, right-facing triangles), semimajor axis distances (orange circles), and aphelion distances (green, left-facing triangles) of the known outer main-belt MBCs, as well as the range of heliocentric distances over which they have been observed to exhibit activity (thick black horizontal lines).
  }
\label{figure:eqtemp_heliodist}
\end{figure}

Using these equations, we compute and plot the equilibrium surface temperatures of sublimating and non-sublimating grey-bodies as functions of heliocentric distance over the range of the main asteroid belt (Figure~\ref{figure:eqtemp_heliodist}).  
Non-sublimating isothermal bodies span an equilibrium temperature range of $160~{\rm K} < T_{eq} < 200$~K and non-sublimating subsolar temperatures span a range of $220~{\rm K} < T_{eq} < 280$~K over the semimajor axis ranges of the main asteroid belt, with the outer main belt having a temperature range of $160~{\rm K} < T_{eq} < 170$~K for isothermal bodies and $220~{\rm K}< T_{eq}< 235$~K for subsolar surfaces.  The relatively high eccentricities of the known MBCs cause them to experience temperature variations of $\sim$25--45~K over the course of their orbits (i.e., from perihelion to aphelion) in the non-sublimating, isothermal approximation, and up to $\sim$35--65~K in the non-sublimating subsolar approximation.  These large temperature variations are due to the fact that the incident solar flux is $\sim$2--3 times greater at perihelion than at aphelion for these objects, and are associated with large variations in the predicted sublimation rate.

For both $\chi=1$ and $\chi=4$, the deviation between the temperature curves in Figure~\ref{figure:eqtemp_heliodist} for the sublimating and non-sublimating cases increases with decreasing heliocentric distances as ``sublimation cooling'' becomes an increasingly dominant process at higher temperatures.  Isothermal bodies ($\chi=4$) only begin to show significant deviations between the sublimating and non-sublimating cases at $R<2.8$~AU.  This thermal behavior indicates that a negligible fraction of the incident solar flux over the semimajor axis range of the outer main belt and beyond contributes to driving sublimation, and is intriguing because the vast majority of observed MBC activity is also seen to take place at $R<2.8$~AU.

\begin{figure}
\centerline{\includegraphics[width=3.5in]{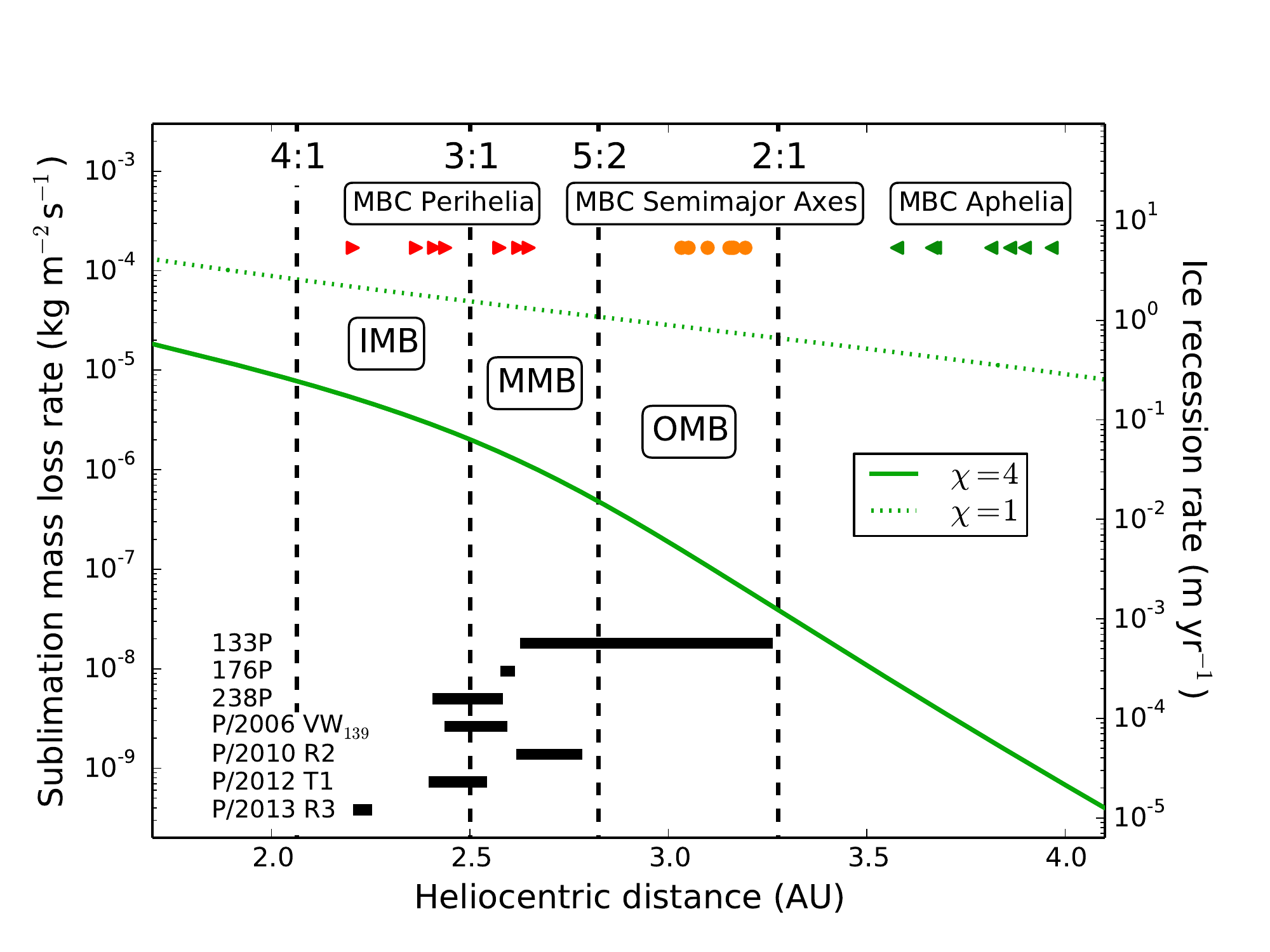}}
\caption{\small Mass loss rate due to water ice sublimation from a sublimating grey body as a function of heliocentric distance over the range of the main asteroid belt, where the semimajor axis ranges of the inner, middle, and outer main belt are labeled IMB, MMB, and OMB, respectively, and mass loss rates calculated using the isothermal approximation ($\chi=4$) and subsolar approximation ($\chi=1$) are marked with solid and dashed green lines, respectively.  The positions of the major mean-motion resonances with Jupiter (4:1, 3:1, 5:2, and 2:1) that delineate the various regions of the main asteroid belt are shown with vertical dashed black lines.  Also plotted are the perihelion distances (red, right-facing triangles), semimajor axis distances (orange circles), and aphelion distances (green, left-facing triangles) of the known outer main-belt MBCs, as well as the range of heliocentric distances over which they have been observed to exhibit activity (thick black horizontal lines).
  }
\label{figure:sublim_rate_heliodist}
\end{figure}

Figure~\ref{figure:sublim_rate_heliodist} shows the steady-state mass loss rate due to water ice sublimation for a sublimating grey-body with the corresponding equivalent ice recession rate on the right ordinate. Using the isothermal approximation, the sublimation rate for a typical MBC is nearly four orders of magnitude larger at perihelion than at aphelion.  The difference between sublimation rates at perihelion versus aphelion is far smaller in the subsolar approximation, but given that real object temperatures will lie somewhere between the isothermal and subsolar approximations (and likely closer to the isothermal approximation than not, as discussed below in Section~\ref{section:mantles}), this means that most MBCs with moderate eccentricities should experience moderate to large variations in sublimation rates along their orbits.

This finding is compatible with the observed variations of MBC activity strength along their orbits discussed above, where peak activity levels are observed near or shortly after perihelion, and activity weakens to undetectable levels at larger heliocentric distances.  Thus, despite the fact that MBCs have much smaller eccentricities than classical Jupiter-family comets or long-period comets, even their modest eccentricities suffice to explain their observed activity modulation patterns along their orbits.

\subsubsection{Mantle formation}
\label{section:mantles}

In addition to changes in sublimation rate at different heliocentric distances, rubble mantle growth is also important for activity modulation.  As volatile material sublimates from the surface of an active object, particles of intermixed inert material (i.e., dust) are either ejected, if they are small enough to be ejected by the local gas drag, or left behind if they are too large.  Over time, an insulating cohesion-free layer of large particles will accumulate to form a so-called rubble or refractory mantle, which ultimately suppresses further sublimation when it grows sufficiently thick \citep[cf.][]{jew96}.

The growth rate of a rubble mantle, $\dot D_M$, can be approximated by
\begin{equation}
{\dot D_M}\sim f_{M} f_D {\dot m_{w}\over \rho}
\label{eq:DM}
\end{equation}
where $f_M$ is the fraction of solid mass that cannot be ejected by gas drag, and $f_D$ describes the reduction in sublimation efficiency caused by the diffusion barrier presented by the growing rubble mantle (where $f_D=1$ in the absence of a mantle). We can approximate $f_M$ using
\begin{equation}
f_M = {\ln(a_+/a_c)\over\ln(a_+/a_-)}
\end{equation}
where $a_-$ and $a_+$ are the minimum and maximum radii of dust particles present in the surface regolith of an object on which sublimation-driven dust ejection is taking place, and the critical radius, $a_c$, at which a particle can no longer be ejected by gas drag is estimated using
\begin{equation}
a_c\sim{9C_D v_g\over 16\pi G\rho_d\rho_n r_n}\dot m_{w}\approx794.2{\dot m_{w}\over R^{1/2}}
\label{eq:ac}
\end{equation}
where $G$ is the gravitational constant, and we assume a drag coefficient of $C_D\sim1$, an outgassing velocity of $v_g\approx 500 R^{-1/2}$ as a function of the heliocentric distance, $R$, in AU, and $r_n=1$~km for the radius of the object \citep[cf.][]{jew02}.  We also assume $\rho_d=\rho_n=1300$~kg~m$^{-3}$ for the bulk densities of the ejected particles and the object nucleus, respectively, based on a rotation analysis of 133P in which $\rho=1300$~kg~m$^{-3}$ was determined to be the minimum critical density that a gravitationally-bound aggregate (i.e., a ``rubble pile'') with the shape and spin rate of 133P is required to possess in order to remain gravitationally bound against rotational disruption \citep{hsi04}.

Meanwhile, the reduction in the sublimation rate due to the diffusion barrier, $f_D$, is determined by the ratio of pore size to layer thickness \citep{sch08}.  If the effective pore size is crudely estimated as the geometric mean of $a_+$ and $a_c$, then 
\begin{equation}
f_D = {1 \over 1+ {D_M/\sqrt{a_+ a_c}}}.
\label{eq:fD}
\end{equation} 

The thermal skin depth, $L$, of an object is given by
\begin{equation}
L = {I\over\rho c} \left( {P\over\pi}\right)^{1/2}
\end{equation}
where the thermal inertia, $I$, is defined by
\begin{equation}
I = (k\rho c)^{1/2}
\end{equation}
where $k$ is the thermal conductivity, $\rho$ is the bulk density, $c$ is the heat capacity of the regolith material, and $P$ is the time period of temperature oscillations (the orbit period, $P_{orb}$, or the rotation period, $P_{rot}$).  

Assuming $c=500$~J~kg$^{-1}$~K$^{-1}$ and $k=10^{-2}$~W~m$^{-1}$~K$^{-1}$ \citep[cf.][]{hei91}, the diurnal thermal skin depth is $L_{d}\sim1\times10^{-2}$~m for an object with a rotational period of $P_{rot}=6$~hr, meaning that material below this depth from the surface is nearly insulated from diurnal temperature variations, and can be considered to be at the equilibrium temperature for that heliocentric distance.  We also compute a seasonal thermal skin depth of $L_{s}\sim0.9$~m for an object with a typical MBC orbital period of 5.5~yr (cf.\ Table~\ref{table:knownaas}), meaning that material below this depth is nearly insulated from annual temperature variations, and can be considered to be at the average temperature for the object over its entire orbit, or approximately the equilibrium temperature at a heliocentric distance equal to the object's semimajor axis.

\setlength{\tabcolsep}{3pt}
\begin{table}[ht]
\caption{Rubble mantle growth on main-belt comets}
\smallskip
\scriptsize
\begin{tabular}{lcccccccccllrrr}
\hline\hline 
 \multicolumn{1}{c}{} && \multicolumn{5}{c}{$\chi=1$, $D_M=0^a$} && \multicolumn{5}{c}{$\chi=4$, $D_M=L_s^b$} \\
 $R^c$ &&
 $T^d$ & {$\dot m_{w}^e$} & {$a_c^f$} & \multicolumn{1}{c}{$f_D^g$} & \multicolumn{1}{c}{${\dot D_M}^h$} &&
 $T^d$ & {$\dot m_{w}^e$} & {$a_c^f$} & \multicolumn{1}{c}{$f_D^g$} & \multicolumn{1}{c}{${\dot D_M}^h$} \\ 
\hline
 2.5 & & 191 & 4.9$\times$10$^{-5}$ & 2$\times$10$^{-2}$ & 1.0 & 3$\times$10$^{-1}$ && 175 & 2.7$\times$10$^{-6}$ & 1$\times$10$^{-3}$ & 0.54  & 3$\times$10$^{-2}$ \\
 3.1 & & 187 & 2.6$\times$10$^{-5}$ & 1$\times$10$^{-2}$ & 1.0 & 2$\times$10$^{-1}$ && 160 & 1.1$\times$10$^{-7}$ & 5$\times$10$^{-5}$ & 0.19  & 7$\times$10$^{-4}$ \\
 3.8 & & 182 & 1.2$\times$10$^{-5}$ & 5$\times$10$^{-3}$ & 1.0 & 1$\times$10$^{-1}$ && 145 & 2.0$\times$10$^{-9}$ & 8$\times$10$^{-7}$ & 0.03 & 3$\times$10$^{-6}$ \\
\hline
\hline
\end{tabular}
\newline {$^a$ Rubble mantle growth parameters for the subsolar approximation ($\chi=1$) for freshly exposed ice (i.e., no initial mantle)}
\newline {$^b$ Rubble mantle growth parameters for the isothermal approximation ($\chi=4$) for a mantle thickness, $D_M$, equals the diurnal thermal skin depth, $L_s$}
\newline {$^c$ Heliocentric distance, in AU}
\newline {$^d$ Equilibrium temperature with sublimation, in K}
\newline {$^e$ Mass loss rate due to sublimation for surface ice, eq.\ (\ref{equation:sublim3}), in kg~m$^{-2}$~s$^{-1}$}
\newline {$^f$ Critical radius at which particles can no longer be ejected by gas drag, eq.\ (\ref{eq:ac}), in m}
\newline {$^g$ Sublimation efficiency, eq.\ (\ref{eq:fD})}
\newline {$^h$ Mantle growth rate, eq.\ (\ref{eq:DM}), in mm/day}
\label{table:rubblemantles}
\end{table}

For freshly exposed ice, the subsolar approximation ($\chi=1$) may be appropriate.  However, once a diurnally insulating rubble mantle has formed, material located below the diurnal thermal skin depth effectively assumes the diurnally averaged temperature at the object's current heliocentric distance, making the isothermal case ($\chi=4$) the better approximation.  As the mantle grows thicker, subsurface ice experiences lower peak temperatures, the diffusion barrier becomes thicker, and the reduced gas flow is less able to eject large particles, which leads to a more fine-grained and thus less permeable mantle.  All of these factors decrease the water emission rate.  For our analysis here where we are concerned with observable activity, we are interested in the timescale for the formation of a mantle that is thinner than the seasonal skin depth, but may be several diurnal skin depths thick.

Assuming $a_-=10^{-8}$~m and $a_+=10^{-1}$~m, we compute water sublimation rates (for surface ice) and critical dust particle radii for objects at typical perihelion, semimajor axis, and aphelion distances for outer-belt MBCs (2.5~AU, 3.1~AU, and 3.8~AU, respectively) for these two cases, i.e., no mantle and $\chi=1$, and a mantle with a thickness on the order of the diurnal skin depth and $\chi=4$ (Table~\ref{table:rubblemantles}).  We also compute sublimation efficiencies, effective sublimation rates (i.e., the product of the base sublimation rate and the sublimation efficiency), and mantle growth rates in these two situations.

Our results show that in the absence of a mantle (i.e., for fresh surface ice), where we assume $\chi=1$, the effective sublimation rate and mantle formation rate are initially high at any position in the orbit of a typical MBC (from $R=2.5$~AU to $R=3.8$~AU).  As the subsolar regime transitions into the isothermal regime, the effective sublimation rate and the mantling rate decrease, and also become more strongly dependent on heliocentric distance, decreasing by about four orders of magnitude from perihelion to aphelion for $\chi=4$ (cf.\ Figure~\ref{figure:sublim_rate_heliodist}; Table~\ref{table:rubblemantles}).  During this transition, strong activity, and therefore significant mantle growth (since the mantle growth rate is directly related to activity strength), should increasingly become confined to the region of the MBC's orbit near perihelion.  A detailed consideration of the transition between the subsolar and isothermal regimes is beyond the needs of our analysis here, as it is sufficient to simply note that the formation timescale of a diurnally insulating mantle should lie between the formation timescales calculated under each approximation.  A simple iterative calculation indicates that under the subsolar approximation, the mantle should reach one diurnal skin depth in thickness in a timescale of $\tau_{L_s}\sim0.1$~yr (i.e., several weeks) from the time of the initiation of sublimation, while using the isothermal approximation, we calculate that the mantle should reach one diurnal skin depth in thickness in $\tau_{L_s}\sim1.1$~yr.  Therefore, the true formation time of a diurnally insulating mantle should lie in the range of $\tau_{L_s}\sim0.1-1$~yr, where we note that the longer formation timescales calculated using the isothermal approximation may actually correspond to the duration of multiple orbit periods since they only include the total time spent close to perihelion.

Eventually, the mantle should become so thick that it reduces even the peak sublimation-driven dust production rate to below detectable levels.  At this point, the object effectively becomes permanently inactive, that is until another impact event occurs and strips away enough surface material for sublimation to begin again.  Given that the timescale for another such impact to occur on the same body is far larger than a single orbit period for a typical MBC, however \citep[cf.][]{hsi09}, the crude estimates in Table~\ref{table:rubblemantles} provide a plausibility check that a MBC exhibiting cometary mass loss due to the sublimation of recently collisionally excavated ice at small heliocentric distances can be repeatedly active over multiple orbit passages \citep[e.g., 133P and 238P;][]{els96,hsi04,hsi09a,hsi10b,hsi11b,hsi13b,jew14b}.

\citet{kos12} have carried out detailed model calculations of activity and mantle formation for comet P/2008 R1 (Garradd), with several sets of example parameters, and found that the emission of water decreases from one orbital period to another, but in some cases only slowly.  \citet{hsi04} and \citet{jew14b} also computed comparable mass loss rates for 133P for active episodes separated by more than a decade (after accounting for different values for grain size and density used in each analysis).  Our crude estimates in the first row of Table~\ref{table:rubblemantles} (i.e., for $R=2.5$~AU), where ice initially retreats rapidly but slows significantly after a mantle of diurnal skin depth has formed, are consistent with these results.

\begin{figure}
\centerline{\includegraphics[width=4.3in]{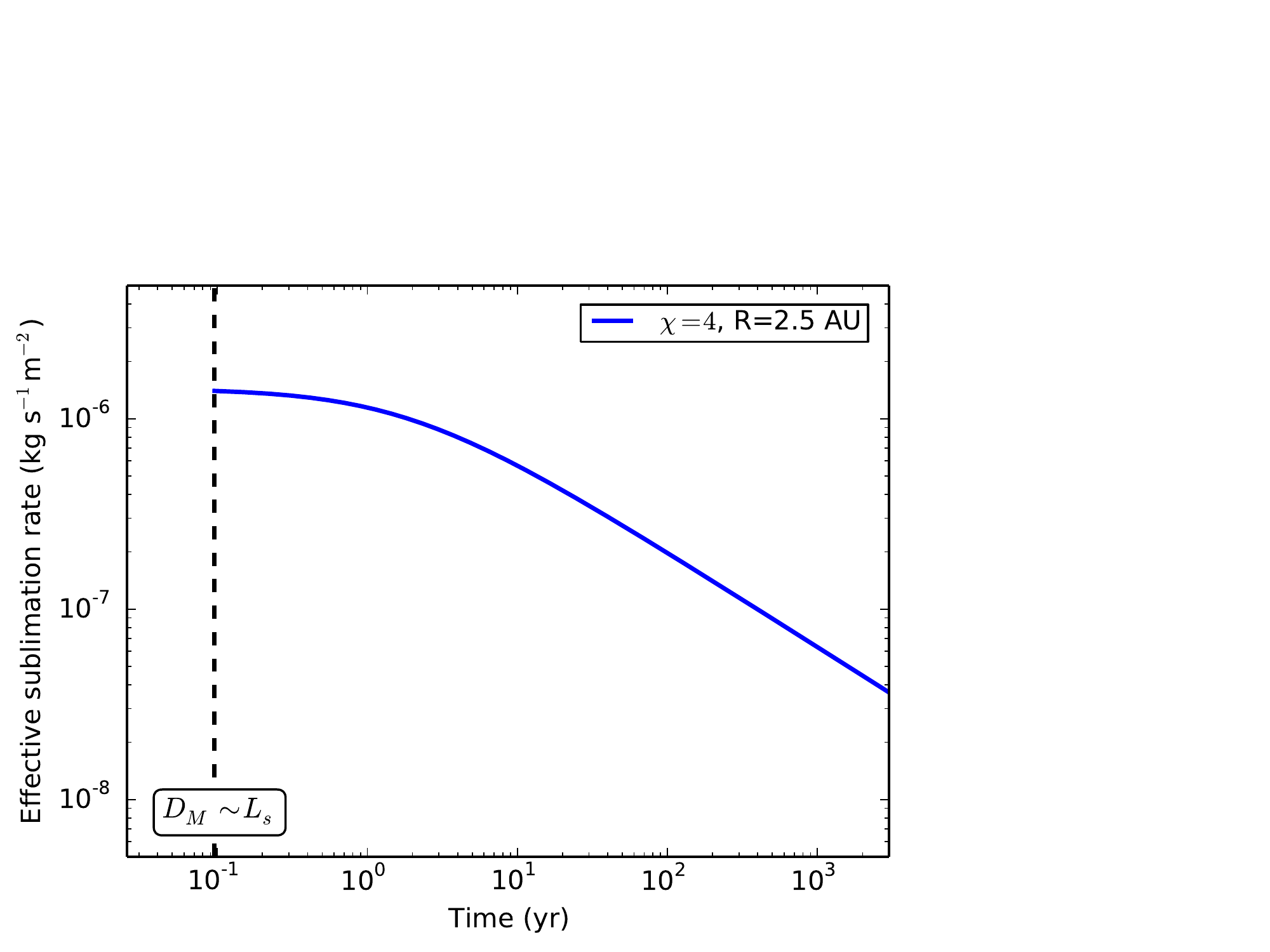}}
\caption{\small Plot of the effective water ice sublimation rate ($f_D{\dot m_w}$) for a typical MBC as a function of time spent near perihelion ($R\sim2.5$~AU) (blue line).  We begin our calculations at 0.1~yr after the start of sublimation, which is the estimated time required for a diurnally insulating rubble mantle to form under the subsolar approximation (vertical dashed line), assume an initial mantle depth of $D_M = L_s$, and also assume $\chi=4$ for the duration of the modeled mantle growth.
  }
\label{figure:mantle_growth}
\end{figure}

Intriguingly, this analysis indicates that there could be a physical basis for expecting widespread low-level activity in the asteroid belt, as suggested by \citet{son11}.  We schematically illustrate the isothermal phase of mantle growth in Figure~\ref{figure:mantle_growth} where we numerically calculate and plot the effective sublimation rate ($f_D{\dot m_w}$) at the perihelion distance of a typical MBC ($R=2.5$~AU) as a function of time spent near perihelion.  The calculations start at 0.1~yr (i.e., the time required for a diurnally insulating rubble mantle to form under the subsolar approximation; vertical dashed line in Figure~\ref{figure:mantle_growth}) after the start of sublimation, assume an initial mantle depth of $D_M=L_s$, and assume $\chi=4$ for the duration of the modeled mantle growth.  Activity is seen to decline asymptotically, becoming suppressed to half of its initial strength on a timescale of $\sim5$ years of mantle formation time, and then taking another $\sim25$ years to halve again.

This asymptotic decline in activity strength means that a MBC should spend a long period of time in a weakly active state, where we assume that the currently known MBCs exhibit comparatively strong activity (since strong activity is easier to discover than weaker activity).  This means that there could be a large population of MBCs that formerly exhibited mass loss rates comparable to that of the currently known objects \citep[$\sim0.1-1$~kg~s$^{-1}$; e.g.,][]{hsi09a,hsi11a,mor11a,mor13,lic13a,jew14b}, but now exhibit long-lived but weaker activity that is just beyond the detection limits of current surveys.  More sensitive future surveys capable of detecting weaker cometary activity than current surveys could be able to access this segment of the active MBC population, and we might therefore expect many more MBC discoveries to result from future search efforts.  Re-computation of mass loss rates from observations of 133P during its many observed active episodes \citep{els96,hsi04,hsi13b,jew07} using consistent values for relevant physical parameters and identical analysis techniques could also help test this hypothesis and set constraints on the rate of MBC activity attenuation.

For simplicity, we have assumed that the water sublimation rate is directly related to the dust mass loss rate.  The dust to gas ratio in MBCs is unconstrained by observations, however, and so a direct relation between the water sublimation rate and dust emission rate may not accurately represent reality.  We also note that rapid rotation or jet-like emission could give rise to larger critical radii than those estimated in Table~\ref{table:rubblemantles} \citep[cf.][]{jew02,hsi11a}, leading to slower mantle growth.  Nonetheless, the timescale for the growth of a seasonal insulating rubble mantle, even in the extremely limiting case of a typical MBC at aphelion in the isothermal approximation, is still extremely short compared to the estimated dynamical lifetimes of these objects \citep[from $\sim$20~Myr to $>2$~Gyr;][]{jew09a,hag09,hsi12b,hsi12c,hsi13a}.  As such, they cannot have been continuously active over their entire residence times in the main asteroid belt, and therefore a recent event, such as an impact that excavated subsurface ice, must have occurred on each of these objects to enable observable present-day sublimation to take place.  

\subsubsection{Discussion}
\label{section:conclusionsthermal}

The mechanism that we explore here, where the thermal wave associated with the perihelion passage of a MBC penetrates through a dry mantle until it reaches the ice and activates the comet via sublimation, is consistent with the pattern of nearly all MBCs exhibiting activity close to or shortly after perihelion.  Since the heat wave penetrates only slowly through the layer, there is a phase delay between the surface temperature and the temperature experienced by the ice.  The thickness of the refractory mantle, relative to the thermal skin depth, can be expected to be associated with a delay in the activation \citep[e.g.,][]{hsi11b}.   Without a mantle, activity may be expected to be nearly symmetric around perihelion.  With a thick mantle, activation can be delayed even until after perihelion is reached.

Bodies in the middle main belt are typically warmer than in the outer main belt, and thus less likely to retain ice near the surface.  This may not be the only reason why active MBCs are predominantly observed in the outer main belt and rarely in the middle main belt though.  A population of middle-belt MBCs could exist, and upon being collisionally activated, the activity of these objects could be much more vigorous than for outer belt MBCs.  However, this activity would also be expected to be shorter-lived due to more rapid mantle formation, and therefore less likely to be discovered than activity in outer-belt MBCs.

The aim of this discussion is merely to outline a possible framework for understanding the observed pattern of MBC behavior, where more work is needed to examine other aspects of the activity modulation mechanism described here.   For example, \citet{sch08} showed that an object's orbital obliquity can significantly reduce the overall average temperature of its surface over its lifetime, meaning that for objects with non-zero obliquity, the temperatures and sublimation rates discussed above represent upper limits to realistic cases.  Moreover, mantles of very low thermal conductivity cause additional cooling due to the amplitude dependence of $T^4$ thermal emission. A full treatment of this problem via detailed numerical modeling, which is beyond the scope of this work, should also include a more systematic exploration of the ranges of possible rotation rates, densities, porosities, heat capacities, thermal diffusivities, thermal inertias, ice-to-dust ratios, total volatile contents, and particle size distributions than are considered here.  The analysis here is intended only to illustrate a possible mechanism for MBC activity modulation using physically plausible parameters, while a more complete exploration of the available parameter space would better determine the range of material properties and environmental conditions under which MBC activity can plausibly arise.  In this regard, the discovery of even more MBCs as well as continued observational studies of known MBCs to determine their physical properties (e.g., nucleus sizes and rotation rates) will also help constrain various aspects of the model framework outlined here.

\section{SUMMARY \& CONCLUSIONS}
\label{section:summary}

We report the following key points and make the following conclusions:
\begin{enumerate}
\item{We analyze a set of 760\,475 observations with $5<S/N<125$ of 333\,026 unique main-belt objects obtained by the PS1 survey telescope between 2012 May 20 and 2013 November 9 using $g_{\rm P1}$, $r_{\rm P1}$, $i_{\rm P1}$, or $w_{\rm P1}$ filters.  These observations include the discoveries of two main-belt comets, one disrupted asteroid, one active Centaur, eighteen Jupiter-family comets, and six long-period comets, as well as several observations of known main-belt comets, disrupted asteroids, and other comets.}
\item{The comet detection procedures currently in use for the PS1 survey consist of the comparison of the point-spread-functions of moving objects to those of reference stars in the same field, and the flagging of objects (on the order of several hundred to several thousand per night) that show significant PSF excesses for human inspection (details in \ref{appendix:comet_screening}).  The majority of these flagged detections can be immediately rejected as false detections, while a small number are selected for manual PSF analysis, and an even smaller number are selected for observational follow-up.  Known issues that can result in missed comet discoveries include the misidentification of real objects as false detections, the subjectivity of the human review process, and the dependence of the discovery process on successful follow-up observations.  Morphology parameter measurements are also known to be less reliable for low S/N detections or detections in close proximity to bright field stars or galaxies, chip gaps, other detector defects, or masked regions, causing a further decrease in our comet discovery efficiency.  Based on the number of missed discovery opportunities by PS1 among comets discovered by other observers, we estimate an upper limit discovery efficiency rate of $\sim$70\% for PS1.}
\item{Based on the statistics of the PS1-observed sample of asteroids observed shortly after perihelion ($0^{\circ}<\nu<45^{\circ}$) where we expect activity detectable by PS1 to be present, we find an expected fraction of $f_{50}=59$~MBCs per $10^6$ outer main-belt asteroids and a 95\% confidence limit upper limit of $f_{95}=96$~MBCs per $10^6$ outer main-belt asteroids, assuming a detection efficiency rate of $C=0.5$, corresponding to a total expected population of $\sim140$~MBCs and an upper limit population of $\sim230$~MBCs with absolute magnitudes of $12<H_V<19.5$ and activity levels detectable by PS1 (i.e., equivalent to mass loss rates on the order of $\sim0.1-1$~kg~s$^{-1}$).}
\item{The known MBC population in the outer main asteroid belt has significantly higher eccentricities than the background asteroid population.  At these eccentricities, the theoretical sublimation rate at perihelion is orders of magnitude larger than at aphelion, consistent with observations showing that MBCs peak in activity strength near perihelion.  This implies that MBC activity is predominantly modulated by variations in heliocentric distance, rather than seasonal variations in solar illumination of isolated active sites related to the orientation of the rotation axis.  These results indicate that the overall mantle growth rate should be slow, as significant mantling should only occur near perihelion, pointing to a process by which MBC activity can be sustained over multiple orbit passages.}
\item{We review lessons learned from the PS1 survey that may help to improve the comet detection capabilities of future surveys (\ref{appendix:futurelessons}).  At the current time, PS1 is unable to detect or is poor at detecting extremely bright objects exhibiting relatively faint activity (i.e., Scheila-type objects), low-level activity with little or no effect on an object's PSF (i.e., 133P-type objects), unresolved activity of the type that could be detected photometrically (i.e., Chiron-type objects), and outbursts on extremely bright and large known comets (i.e., 17P/Holmes-type objects).  Implementation of multiple screening methods, including radial PSF comparison, linear PSF comparison, modeling of trailed PSFs, detection of azimuthally localized activity, comparison of photometry optimized for point sources and photometry optimized for extended sources, detection of photometric deviations from expected brightnesses, and crowd-sourcing, should help to fill these blind spots in future surveys.  Regardless of what method or methods are used, however, we emphasize that false detection minimization, balance between screening sensitivity and human and computational overheads, and robust follow-up observation plans should also be important considerations in the formulation of future comet detection systems.}
\item{PS1's exceptional success at discovering MBCs is likely largely due to its sensitivity to faint, low-activity comets. Future surveys that will be sensitive to even fainter and lower-activity comets should find many more MBCs, perhaps even in excess of the expected population estimated in this paper, and therefore the development of robust comet detection algorithms for these surveys should be a high priority.  Sensitivity to faint comets will also aid studies of active Centaurs, and long-term studies of Earth-approaching comets.
}
\end{enumerate}

The recent detection of water vapor emission from Ceres \citep{kup14} and probable detections of water ice on main-belt object (24) Themis \citep{riv10,cam10} have provided additional evidence of present day water in the asteroid belt.  These findings support the notion that water ice has been able to survive in a number of main-belt asteroids until the present day, and that MBC activity is being driven by sublimation.  Two priorities for future MBC research are the discovery of more MBCs, and the continued physical and dynamical characterization of individual MBCs and the population in general.  While PS1 has been exceptionally successful at discovering MBCs (and disrupted asteroids) compared to previous and ongoing surveys, a substantial increase in the discovery rate is still needed to enable statistically meaningful analyses to be conducted of the spatial distribution of these objects.  In turn, such analyses will enable the use of MBCs to trace the current water content of the asteroid belt, and therefore infer its primordial water content and distribution, subject of course to various dynamical disturbances that very likely occurred between the formation of the solar system and the present day \citep[e.g.,][]{wal11,dem14}.  Fortunately, the discovery power to accomplish such a leap should soon be available with wide-field high-spatial-resolution surveys using large aperture telescopes like the HSC survey and LSST.  By going deeper than any previous surveys before them, they should be able to discover many more new MBCs, shedding new light on these still-enigmatic objects, and perhaps on our own origins as well.

\section*{Acknowledgements}
We thank Pedro Lacerda and Bin Yang and two anonymous referees for helpful comments on this manuscript. 
H.H.H.\ acknowledges support for this work by the National Aeronautics and Space Administration (NASA) through Hubble Fellowship grant HF-51274.01 awarded by the Space Telescope Science Institute, which is operated by the Association of Universities for Research in Astronomy (AURA) for NASA, under contract NAS 5-26555.
N.S. acknowledges support by NASA through the NASA Astrobiology Institute under Cooperative Agreement No.\ NNA09DA77A issued through the Office of Space Science.
J.K. acknowledges support through NSF grant AST 1010059.
The Pan-STARRS1 Surveys (PS1) have been made possible through contributions of the Institute for Astronomy, the University of Hawaii, the Pan-STARRS Project Office, the Max-Planck Society and its participating institutes, the Max Planck Institute for Astronomy, Heidelberg and the Max Planck Institute for Extraterrestrial Physics, Garching, The Johns Hopkins University, Durham University, the University of Edinburgh, Queen's University Belfast, the Harvard-Smithsonian Center for Astrophysics, the Las Cumbres Observatory Global Telescope Network Incorporated, the National Central University of Taiwan, the Space Telescope Science Institute, the National Aeronautics and Space Administration under Grant No.\ NNX08AR22G issued through the Planetary Science Division of the NASA Science Mission Directorate, the National Science Foundation under Grant No.\ AST-1238877, the University of Maryland, and Eotvos Lorand University (ELTE).  We thank the PS1 Builders and PS1 operations staff for construction and operation of the PS1 system and access to the data products provided. 

\appendix

\section{PS1 Comet Screening Procedure Details\label{appendix:comet_screening}}

In terms of the individual steps required for an object observed by PS1 to be discovered as a comet, first, a minimum of two detections of an object must be made, identified as transients by IPP, and linked as a tracklet by MOPS.  A common situation in which this process may not occur successfully is when observations of a particular object in an observing sequence are lost to chip gaps or other unusable regions of the PS1 camera due to the the camera's $\sim$75\% effective fill factor (which includes the effects of chip gaps, guide star cells, and masking of bad pixels, detector artifacts, and diffraction spikes).  Tracklet linking can also fail if variable sky conditions hinder successful subtraction of background sources, or cause some images of faint objects to fall below IPP's detectability threshold within an observing sequence even if the object is detectable in other exposures.  If tracklet linking is completed successfully, the tracklet must then be flagged by the MOPS comet screening system, but as discussed above, reliable morphology parameter measurements (on which our flagging system relies) from IPP are not guaranteed, particularly for low $S/N$ detections or detections in close proximity to bright field stars or galaxies, chip gaps, or masked regions.

A notable example of a comet being missed due to issues with tracklet linking is MBC P/La Sagra. Following its discovery on 2010 September 14 \citep{nom10}, it was found to have been clearly cometary in two sets of PS1 observations obtained in the month prior to its official discovery \citep{hsi12c}.  On one night, it was observed twice, and in another, it was observed four times.  However, the high false detection rate at the time meant that pairs of detections were considered unreliable, and so the first set of observations was simply ignored by the system.  In the set of four detections, the object passed near a bright field star in one image, skewing its measured magnitude.  At the time, MOPS required detections within a tracklet to have similar magnitudes to ward against mis-linked detections (this is no longer true, largely due to this case), and so this tracklet was rejected and also never displayed for human review.

\begin{figure}[t]
\centerline{\includegraphics[width=4.3in]{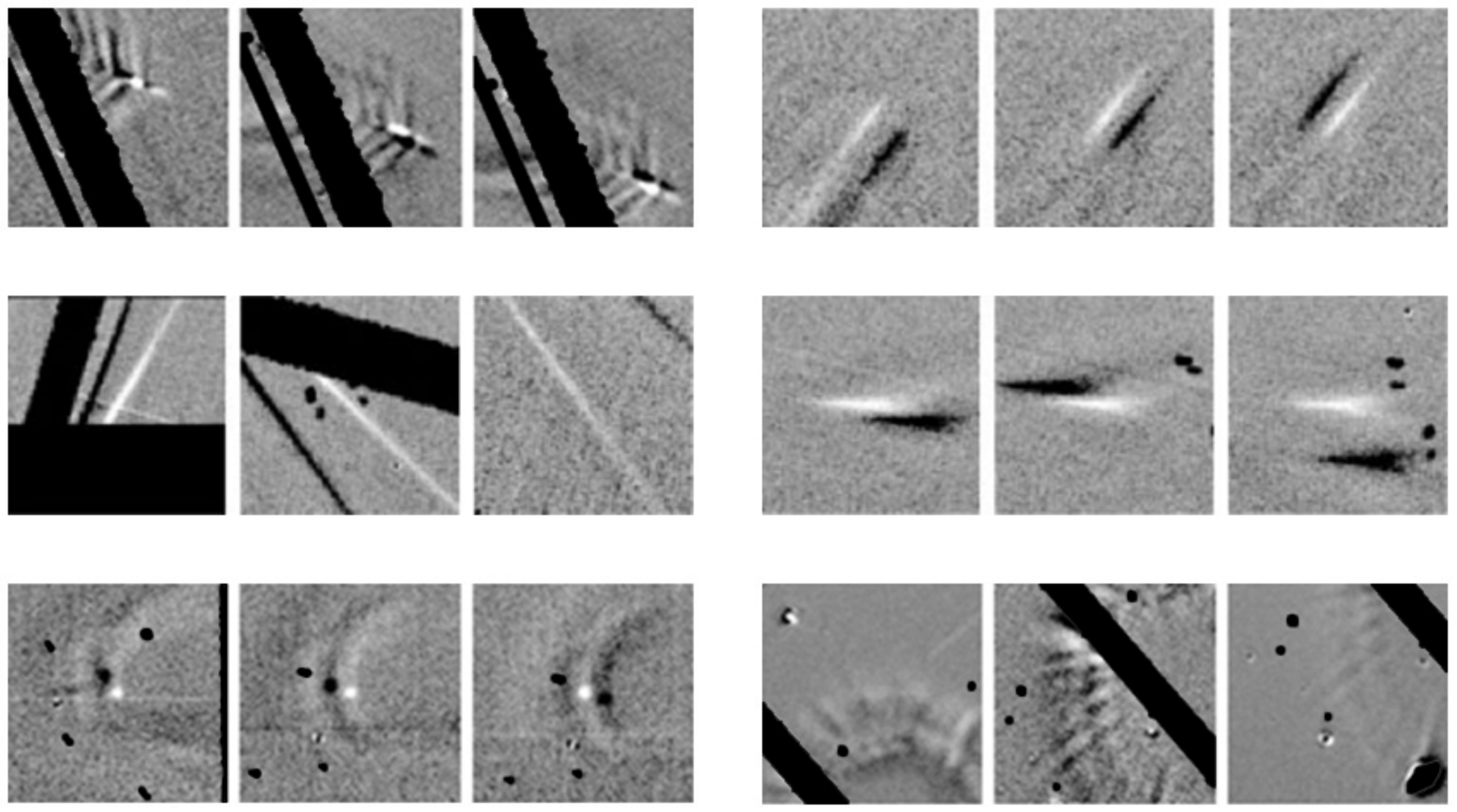}}
\caption{\small Difference images of representative false detections from PS1.
  }
\label{figure:falsedetections}
\end{figure}

The automated portion of the PS1 comet screening process identifies several hundred to several thousand comet candidates each night.  Of these, most are data artifacts (e.g., internal reflections, imperfectly-subtracted stationary sources, and diffraction spikes; Figure~\ref{figure:falsedetections}), low $S/N$ detections, or saturated detections, giving a relatively large false positive rate on the order of a few hundred to one.  It is however at a level that can be managed by a secondary level of human screening, where real objects can be rapidly identified by eye by a MOPS team member who then makes a visual assessment and, if warranted, a quantitative assessment of the cometary nature of the object.  This initial evaluation process is subjective and can depend on the experience of the human screener on a given day, the amount of time that can be devoted to each day's screening given other operational priorities, and the number of objects flagged for evaluation on any particular night.  This process can also be affected by other candidates in the same data set, since the discovery of several good candidates in a single night could necessitate the prioritization of follow-up efforts at the expense of lower-probability candidates, while a data set of generally poor quality may not be reviewed as carefully as a better quality data set.

Comet candidates that pass all the screening stages up to this point must then be scheduled for follow-up observations to confirm the activity, and also extend the orbital arc if the candidate is an unknown object.  Follow-up observations by PS1 team members are typically conducted with the University of Hawaii 2.2~m telescope or CFHT, both on Mauna Kea in Hawaii, or the 2.0~m Faulkes Telescopes (North and South) on Haleakala in Hawaii and Siding Spring in Australia.  However, scheduling, weather, instrument availability, or other technical issues can sometimes prevent us from using these facilities.  External observers are sometimes able to provide follow-up confirmation and astrometry, since we report all of our high-probability comet candidates to the MPC upon identification.  Prior to October 2013, the MPC would place comet candidates on their Near-Earth Object Confirmation Page where experienced amateur observers and professional astronomers can view a list of high-priority targets needing follow-up observations.  Starting in October 2013, the MPC introduced a new Possible Comet Confirmation Page, where comet candidates needing confirmation are now listed.  However, the quality of observing facilities available to these external observers varies widely, and for very difficult (i.e., faint or marginal) comet candidates, sometimes follow-up on 2-m-class professional telescopes or larger is required to obtain the observations that are needed.

If an object is already known and confirmation of cometary activity is all that is needed, delays in obtaining follow-up observations are often tolerable as cometary activity typically remains observable for at least several days, or even several weeks or months.  Since the time baseline spanned by the typical tracklet is usually only about one hour though, positional uncertainties based on initial orbit determinations of previously unknown objects are often large and increase rapidly.  As such, follow-up typically has to be performed within several days of the original PS1 observations before the positional uncertainties simply grow too large for the object to be recovered, even by as large a camera as CFHT's MegaCam and its $1^{\circ}\times1^{\circ}$ field of view, in which case, the object is effectively lost.  Occasionally, precoveries or self-recoveries of objects are found among other PS1 observations, but due to survey constraints, we cannot perform targeted follow-up with the PS1 telescope itself.

PS1 observations of moving objects are not currently automatically stacked to create higher $S/N$ composite images (although they are often manually stacked during the human-screening stage of the comet identification process).  While such composite images are normally useful for searching for low-surface-brightness activity, the current level of inconsistency between PS1 observations of moving objects is sufficiently high that this technique would have limited usefulness as part of the automatic screening process.  The small sizes of the individual CCDs in the PS1 detector array means that one or more detections in a tracklet may be obstructed by a chip gap, thus contaminating the remaining gap-free images if these obstructed detections are indiscriminately included when stacking.  Chip defects that are automatically masked by IPP present a similar issue.  In most cases, median stacking is unhelpful for compensating for this problem due to the small number of detections per tracklet.  Stacking data also hinders visual identification of poorly subtracted background objects that could be mistakenly interpreted as faint cometary features.

Given these considerations, we currently prefer to consider individual detections, looking for consistency of morphology measurements and visual appearance of detections within each tracklet.  Promising detections flagged during visual screening can then be manually stacked, allowing any undesirable detections within a tracklet to be identified and omitted.  Given the added sensitivity to low-level cometary activity that would be gained, though, automated stacking of all tracklets may be worthwhile to perform during post-processing of PS1 data, as well as in real-time for future surveys, if the complications discussed here can be mitigated.

\section{Lessons for Future Surveys\label{appendix:futurelessons}}

Despite PS1's success to date at discovering a wide range of comet-like objects, from classical comets to MBCs to DAs, it is important to note what types of comet-like objects PS1 is currently unable to detect, i.e., the system's ``blind spots'', so that future surveys can attempt to fill similar gaps in their comet discovery procedures.  For example, comparison of asteroidal and stellar radial PSFs is effective for detecting coma in minimally trailed objects, as we have demonstrated over the course of the PS1 survey, but is less effective for objects that are significantly trailed (i.e., for objects with large non-sidereal velocities, long exposure times, or both).  This complication can be handled by measuring linear PSFs perpendicular to the direction of trailing for both asteroids and stars instead of measuring radial PSFs.  This method will be sensitive to spherical coma or directed emission perpendicular to the direction of trailing, and will be insensitive to directed emission parallel to the direction of trailing, but at least permits trailed objects to be searched for activity at all.  This type of analysis is already often manually performed on candidate comet images in PS1 data as part of the human review process to determine whether detected PSF excesses are due to cometary activity or trailing effects, but automating the process for future surveys could prove productive.

PS1 is currently unable to detect Scheila-type objects, i.e., large, bright objects that undergo substantial outbursts due to impacts or cometary activity, because such objects saturate the PS1 camera.  This will be an even more severe problem with upcoming wide-field surveys using even larger aperture telescopes like LSST.  This problem could be alleviated by the development of algorithms to detect excess sky flux far from the photocenter of any saturated moving objects, i.e., in the presumably unsaturated wings of those objects' PSFs.  Meanwhile, relatively shallow surveys such as ATLAS \citep{ton13} could also potentially allow bright asteroids to be productively searched for activity.   The potential for confusion of background objects or noise artifacts with comet-like activity by automated routines will be even higher for activity searches at large sky-plane separations from saturated objects, however, compared to searches for close-in activity of unsaturated objects similar to what we describe here.  Thus, even more care will need to be taken to mitigate the effect of false detections in this type of search.

PS1's automated flagging system relies entirely on the measurement of deviations in an object's PSF as compared to PSFs of nearby field stars (Section~\ref{section:ps1comets}), and so another current blind spot for PS1 is the inability to detect comet-like activity (e.g., a faint dust tail) that has minimal effect on an object's PSF.  A notable example of such an object is 133P \citep{hsi04,hsi10b}.  To account for objects like this, we had originally intended to employ a tail-finding algorithm similar to that described in \citet{son11} where a circular annulus is placed around each candidate object and divided into slices.  The flux inside each slice can then be measured, with any excess flux along a particular azimuthal direction noted as an indication of possible directed dust emission such as a dust tail.  This technique unfortunately was only minimally effective for finding comets in PS1 data due to the large number of data artifacts which led to an unmanageably large number of false positive detections, but may be more effective in analyzing higher quality data from future surveys.

Searching for unresolved activity by looking for asteroids exhibiting photometric excesses compared to their expected brightnesses is another method for detecting cometary activity \citep[e.g.,][]{tho88,bus88,cik14} that is currently unimplemented as part of PS1's comet screening procedures.  Work is ongoing to develop mechanisms for reliably detecting comets from photometry alone, but challenges are numerous.  Detecting photometric enhancements in asteroids relies on having accurate knowledge of their photometric brightnesses at the time of observation as well as their expected brightnesses based on their absolute magnitudes and phase functions.  PS1 data are now fully photometrically calibrated across the entire sky \citep{sch12,mag13}, but the lack of accurate photometric references across the entire sky during the early stages of the survey impeded efforts to implement photometric activity detection until just recently.  Furthermore, accurate measurements of asteroid phase functions are lacking for a large segment of the main-belt population, meaning that  predictions of the brightnesses of these objects must rely on relatively low-precision MPC data.  Photometric activity searches are still possible given these circumstances, but would only be sensitive to activity causing photometric enhancements in excess of the uncertainties due to imprecisely known predicted magnitudes as well as those due to imprecisely known observed magnitudes.  However, given the large sizes of these uncertainties (at least until recently), any objects showing such large photometric enhancements were probably exhibiting bright enough activity to be discovered via other means.  This situation is greatly improved now, however, as PS1 data can now be considered photometrically reliable as discussed above, and thanks to PS1, we have also amassed baseline photometry for a much larger segment of the main-belt population than was available before and so have been able to calculate phase functions for these asteroids (Vere{\v s} et al., 2014, in prep).  This method for searching for photometric indications of activity should therefore henceforth be much easier to conduct for both PS1 and other surveys, but will still remain limited to detecting activity that causes photometric enhancements in excess of the brightness uncertainties due to an object's unknown rotational state at the time of observations.

\begin{figure}
\centerline{\includegraphics[width=5.0in]{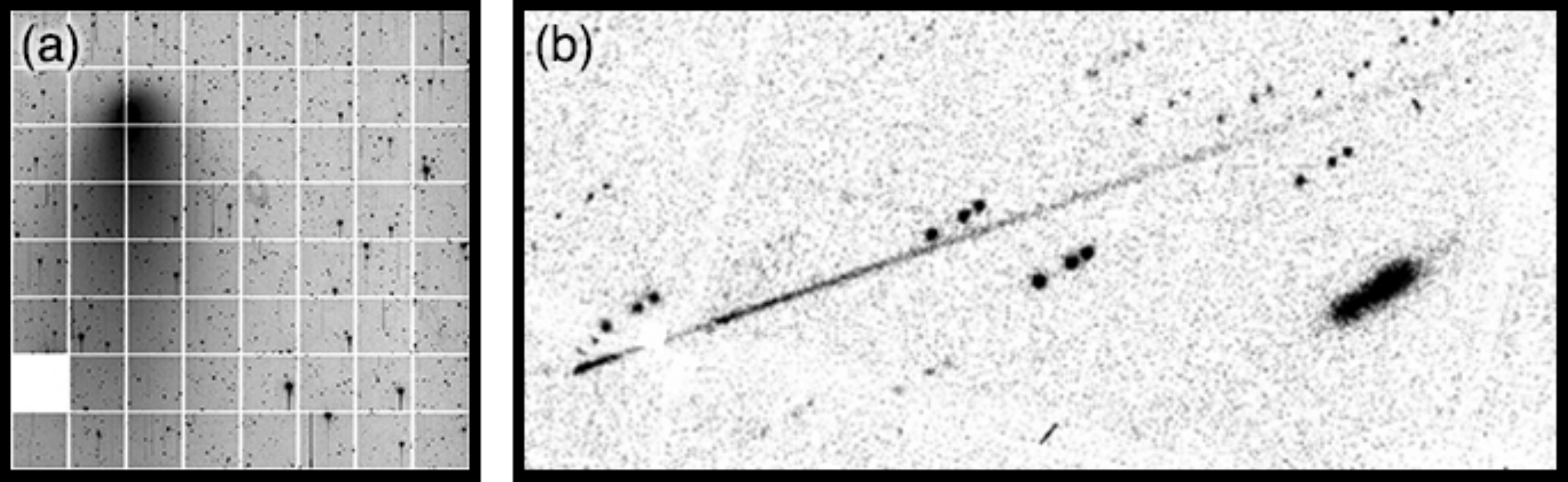}}
\caption{\small (a) $20'\times20'$ image of C/2012 K5 (LINEAR) obtained by PS1 on 2013 January 7. (b) Composite $w_{P1}$-band image of P/Gibbs constructed from data obtained by PS1 on 2011 December 06.  The area of sky shown is $3.5'\times1.5'$ in size with North at the top and East to the left. In this image, the dust trail is observed to extend $\sim4.8'$ to the northwest from the nucleus (lower left corner). The grid-like pattern in the sky is due to chip gaps in the PS1 CCD mosaic.
  }
\label{figure:ps1brightcomet}
\end{figure}

Finally, there is the problem of automating the detection of comets that are extremely bright ($m_V\lesssim14$~mag; cf.\ Table~\ref{table:sampledist}), have large angular sizes, or both (e.g., Figure~\ref{figure:ps1brightcomet}a), because of the difficulty of even identifying them as real sources using PS1's point-source detection software.  Most bright comets are already known and so this problem likely does not result in a large number of missed comet discoveries by PS1. Previously unknown comets have occasionally been discovered due to undergoing sudden outbursts \citep[e.g.,][]{nak10,ish14}, however, and the inability of the PS1 system to automatically detect extremely bright objects means that we also risk missing the discovery of sudden cometary outbursts from known comets such as the one exhibited by 17P/Holmes in 2007 \citep{buz07}.  Comets with unusual morphologies, without distinct central condensations, or that are otherwise extremely non-point-source-like present similar problems.  P/2012 F5 (Gibbs) was discovered on 2012 March 22 \citep{gib12} but appeared visibly cometary in PS1 data as early as 2011 December 6, and again on 2012 February 28 (Figure~\ref{figure:ps1brightcomet}b).  However, because its central condensation was so indistinct compared to the rest of the comet, and the comet itself was dominated by an extremely long and narrow tail, IPP and MOPS filters designed to screen out image artifacts like diffraction spikes rejected these detections, preventing them from ever entering the MOPS comet screening pipeline.

It is difficult to conceive of what reasonable modifications or additions to the PS1 comet detection pipeline would allow us to deal with such unusual objects, and so for finding such comets in future surveys, perhaps the solution lies not in automation, but rather in the outsourcing of comet detection to the public via citizen science projects \citep[e.g.,][]{lin08,smi11}.  Such projects have been used to address problems in astronomy as diverse as classifying galaxy morphologies \citep[e.g.,][]{lin08}, identifying supernovae \citep{smi11}, searching for infrared bubbles in the inner Galactic plane \citep{ken12}, measuring structural properties of galaxies \citep{vin13}, searching for planetary transits in Kepler data \citep[e.g.,][]{fis12,schw12}, and locating precoveries of near-Earth asteroids in SDSS data \citep{sol14}.  Without fixed preconceived notions of what types of comet morphologies to search for that are inherent in any automated detection algorithm, citizen scientists could help ensure that comets with morphologies that are unexpected or simply difficult for automated routines to detect are not missed, while also adding the potential to make serendipitous discoveries of unexpected phenomena \citep[e.g.,][]{car09,lin09}.  While final evaluation of candidate objects would almost certainly still have to be done by professional astronomers, implementing citizen science comet detection programs could be a relatively low-effort way for future surveys to supplement automated detection pipelines with a large-scale human visual screening effort to minimize the chances of missing unusual comets.

Among several other practical lessons we can take from our experience with PS1 that are applicable to future surveys, we find that automated comet detection remains difficult to perform reliably, with major challenges including false detection minimization and coping with the wide range of comet morphologies that are possible.  Unfortunately, at least for PS1, many common data artifacts have very comet-like morphologies (cf.\ Figure~\ref{figure:falsedetections}), and so are routinely flagged as comet candidates by our automated detection system.  Minimizing the number of these obviously false detections was a critical requirement in the development of our comet screening procedures because while a rate of several hundred false detections for every real detection is manageable for a human screener, a rate of several thousand or tens of thousands to one (as we originally had) was simply intractable.  Even then, more subtle false detections frequently still require expert human review to identify, a process that has become comparatively more streamlined over time as we gained experience in recognizing the most common types of subtle false detections, but one that still represents a significant burden on human resources over the course of the survey.  We therefore suggest that development of automated false detection algorithms that are as thorough and reliable as possible should be considered a high priority for any comet discovery efforts as part of future surveys.

In terms of dealing with comets with a wide range of morphologies, the best way to handle this issue may simply be the use of multiple screening methods.  Many possible screening methods that future surveys could implement are discussed above, including radial PSF comparison, comparison of linear PSFs measured perpendicular to the direction of an object's trailing, modeling of trailed PSFs, searching for azimuthally localized excess flux around an object, comparison of an object's total flux to its peak flux as a measure of degree of condensation, searching for photometric enhancements of asteroids of known brightnesses, and crowd-sourcing.  In addition to these screening methods, future surveys could also compare PSF magnitudes (which are measured by fitting a Gaussian point-spread function model to an object, and are optimized for photometry of point-like sources) and so-called model magnitudes (which are optimized for photometry of extended objects such as galaxies, and provide more accurate photometry of resolved sources), where extended objects should show the largest deviations between the two measurements.  This method was employed by \citet{sol10} to find comets in SDSS data, and is also used by SDSS for star-galaxy separation as well \citep{str02}.  Similar screening methods including comparisons of aperture magnitudes and Kron magnitudes to PSF magnitudes were also explored for PS1 comet detection but were ultimately rejected due to decreased reliability of these parameters at faint magnitudes or near chip artifacts such as detector gaps or masks.


The suitability of any of these methods for finding comets will depend on the specific characteristics of a given survey, but also on the cost in either human resources, computational power, or time needed to implement them.  For initial screening, it may only be feasible to perform simple magnitude comparisons or basic PSF comparisons using parameters produced automatically by a survey's general reduction pipeline, rather than more detailed custom analyses requiring access to image data that could represent unacceptable burdens on available computational resources, depending on the amount of data involved and the speed at which the data are processed.  Similarly, analyses that are excessively time-consuming because of the amount of computing resources or human intervention required may risk not being performed at all, and could also adversely affect the timeliness with which observational follow-up of candidates can be scheduled, putting their recoveries at risk.  As such, when choosing the most suitable method or set of methods to search for comets, consideration of overheads is perhaps just as important as consideration of the absolute effectiveness of any particular comet detection scheme.

Finally, we must emphasize that a robust follow-up plan is an essential complement to any survey aiming to find comets.  Follow-up confirmation and astrometry of newly discovered comets has frequently been provided in the past by amateur astronomers using small telescopes ($<0.5$~m) at sub-optimal observing sites.  As comets discovered by advanced surveys include increasingly fainter and weaker comets, either by utilizing increasingly sensitive algorithms (as PS1 does) or increasingly larger telescopes (as the HSC survey and LSST will do), follow-up observations will become increasingly out of the reach of amateur observers, necessitating the use of professional facilities.  Such facilities will generally not need to have comparable aperture sizes to the telescopes conducting the surveys in question, since they can simply use longer integration times.  Ideal requirements for any follow-up facilities, however, do include being located at sites with reliable observing-quality weather, availability of a wide-field imager most or all of the time, and the ability to rapidly respond to observation requests on short notice (e.g., through a queue-scheduled system or a rapid-response target-of-opportunity policy), where we note that self-follow-up by the original survey facility would also be suitable if it can be accommodated as part of ongoing survey operations.

\end{document}